# FiberKAN: Kolmogorov-Arnold Networks for Nonlinear Fiber Optics

Xiaotian Jiang, Min Zhang, Xiao Luo, Zelai Yu, Yiming Meng, Danshi Wang, *Senior Member, IEEE*

*Abstract*— **Scientific discovery and dynamic characterization of the physical system play a critical role in understanding, learning, and modeling the physical phenomena and behaviors in various fields. Although theories and laws of many system dynamics have been derived from rigorous first principles, there are still a considerable number of complex dynamics that have not yet been discovered and characterized, which hinders the progress of science in corresponding fields. To address these challenges, artificial intelligence for science (AI4S) has emerged as a burgeoning research field. In this paper, a Kolmogorov-Arnold Network (KAN)-based AI4S framework named FiberKAN is proposed for scientific discovery and dynamic characterization of nonlinear fiber optics. Unlike the classic multi-layer perceptron (MLP) structure, the trainable and transparent activation functions in KAN make the network have stronger physical interpretability and nonlinear characterization abilities. Multiple KANs are established for fiber-optic system dynamics under various physical effects. Results show that KANs can well discover and characterize the explicit, implicit, and non-analytical solutions under different effects, and achieve better performance than MLPs with the equivalent scale of trainable parameters. Moreover, the effectiveness, computational cost, interactivity, noise resistance, transfer learning ability, and comparison between related algorithms in fiber-optic systems are also studied and analyzed. This work highlights the transformative potential of KAN, establishing it as a pioneering paradigm in AI4S that propels advancements in nonlinear fiber optics, and fosters groundbreaking innovations across a broad spectrum of scientific and engineering disciplines.**

*Index Terms*—**Kolmogorov-Arnold Network, AI for science, scientific discovery, dynamic characterization, nonlinear fiber optics**

## I. INTRODUCTION

NATURAL science is about understanding, mastering, and applying various natural phenomena in the world from the scales of space and time. In the pursuit of exploring natural science, scientific discovery and dynamic characterization of various physical systems have long been the key to advance and transform our understanding of complex physical phenomena and behaviors across scientific and engineering disciplines, which reveals the underlying laws of the systems by establishing mathematical formulas or physical theorems [1]. Traditionally, scientific discovery and dynamic characterization have relied on observation, experimentation, and theories grounded in rigorous first principles, such as conservation laws, fundamental physical principles, and phenomenological behaviors [2]. However, numerous complex physical systems in real-world scenarios remain inadequately explored, where their dynamic behaviors and system solutions cannot yet be quantitatively characterized or satisfactorily explained using conventional approaches. Fortunately, the rapid advancements of artificial intelligence (AI) techniques have paved new avenues for addressing these challenges, and give rise to an emerging research area known as AI for science (AI4S) [3]. Initially, by leveraging multi-layer perceptron (MLP)-based AI algorithms in a data-driven manner, multiple underlying dynamics and physical characteristics in systems previously deemed intractable are discovered, such as protein structure prediction [4], molecular systems modeling [5], and epidemiology analysis [6]. Later, this field considered more of the emerging symbiosis between physics and AI [7], moving towards AI-based physical models [8], physics as computer [9], and physics-inspired algorithms [10]. By enhancing, accelerating, and deepening our understanding of natural phenomena, AI4S will greatly promote the transformation of scientific research paradigms across various fields, thereby advancing humanity's exploration of the unknown problems.

In fiber-optic systems, the evolution of optical pulse is influenced by the varying responses of different dielectrics to light under high-intensity electromagnetic fields, leading to a range of physical effects, including power attenuation, chromatic dispersion, and multiple nonlinear effects [11]. These physical effects collectively give rise to complex dynamic behaviors, which are crucial for discovering uncertain fiber physics [12], designing sophisticated fiber structures [13], and developing advanced optical fiber communications [14]. The fundamental step toward these progresses is the mathematical characterization and physical explanation of the mechanisms behind the various effects. Typically, the propagation dynamics of optical pulses in fiber can be mathematically characterized by the nonlinear Schrödinger equation (NLSE) [11], which is derived from Maxwell's equations based on first principles. To

This work was supported in part by the National Natural Science Foundation of China under Grant 62171053, Beijing Nova Program under Grant 20230484331, and BUPT Excellent Ph.D. Students Foundation under Grant CX2023226. (*Corresponding author: Danshi Wang*).

Xiaotian Jiang, Min Zhang, Xiao Luo, Zelai Yu, Yiming Meng, and Danshi Wang are with the State Key Laboratory of Information Photonics and Optical Communications, Beijing University of Posts and Telecommunications, Beijing 100876, China (e-mail: jxt@bupt.edu.cn; mzhang@bupt.edu.cn; XiaoLuoD@bupt.edu.cn; ytzelai@163.com; 2024140620@bupt.edu.cn; danshi_wang@bupt.edu.cn).

solve these governing equations, numerical methods such as the split-step Fourier method (SSFM), finite difference method, and finite element method are usually required to approximate the solutions of most fiber-optic nonlinear systems. Although SSFM achieves high accuracy, it often necessitates a very small step size in complex scenarios to meet the precision requirements, significantly increasing the computational complexity and limiting the practical applicability. To address these challenges and characterize the underlying dynamics more efficiently, multiple deep learning (DL)-based models have been established to simulate and characterize the physical laws governing fiber-based systems [15-17]. By mapping the system inputs to outputs in a purely data-driven manner, DL-based models aimed to approximate the system's dynamic process without exploring the explicit physical mechanism. In recent years, a physics-informed neural network (PINN) has been proposed [18], providing a promising approach to incorporate physical constraints into network training. In fiber optics, PINNs have also been applied to the fiber dynamics solving [19], fiber channel modeling [20], and fiber parameters identification [21]. By embedding the governing equations and physical constraints (e.g., initial and boundary conditions) into the loss terms, PINNs are able to learn the system dynamics or equation parameters while tuning the network hyperparameters. Such network learning mode greatly reduces the amount of training data, and more importantly, renders the training trajectory of the DL model to be physically consistent.

Although PINNs perform strong potential for characterizing and discovering the dynamics of various complex physical systems, the relationship between network parameters and network outputs remains enigmatic. Similar to most data-driven DL algorithms, the network structures of PINN are typically based on multi-layer perceptron (MLP), also known as fully-connected feedforward neural networks [22]. As the most widely-used foundational building blocks of modern DL models, the value and importance of MLPs will never be exaggerated due to its powerful capability for nonlinear function expression [23]. However, MLP-based structures consistently encounter several challenges in tasks of system dynamics learning: on the one hand, they are typically less interpretable without post-analysis tools, which complicates efforts to fully understand system mechanisms and limits their ability to perform reliable scenario extrapolation [24]; on the other hand, a large number of parameters are required to be optimized for satisfactory performance—for example, MLPs consume almost all non-embedding parameters in transformers [24]. Moreover, MLP-based networks also face the spectral bias phenomenon of preferentially learning low-frequency features [25], thereby leading to great difficulties in learning high-frequency dynamics. Despite some advancements [26-28], these works are not fundamentally distinct from the fully-connected structure, and their performances remains constrained by the constrained by fundamental properties of MLPs. Therefore, an interpretable and concise network structure is desired for scientific discovery and dynamic characterization of complex systems from an AI4S perspective.

In 2024, a Kolmogorov-Arnold Network (KAN) has been proposed [29], garnering significant attention due to its innovative structure, strong interpretability, and great transparency. Unlike MLPs that place fixed activation functions on nodes (i.e., neurons), KAN introduces a paradigm shift by placing learnable activation functions on edges (i.e., weights), which is inspired by the Kolmogorov-Arnold representation theorem [30]. Specifically, each weight parameter in traditional MLPs is replaced by a learnable function parameterized as a Spline in KANs. Simultaneously, outputs from the previous layer are aggregated at the nodes, eliminating the need for nonlinear activation functions. Compared with fixed activation functions, the learnable Splines provide superior nonlinear function approximation capabilities, enabling KANs to achieve higher accuracy with a more compact network structure. Moreover, each Spline function can be clearly visualized, facilitating a more intuitive analysis of the relationship among input, intermediate, and output nodes. This transparency enhances the interpretability, making the network's internal operations more accessible and fostering trust in its predictions. In addition to the Spline function, other basis functions are also applied to construct activation functions for further enhancing the training efficiency or representation ability of KANs [31-37]. Consequently, as a prospective alternative to MLPs, KANs provide a precise, interpretable, and streamlined framework for discovering and characterizing the dynamics of complex systems. By addressing the inherent limitations in MLPs, KANs hold the promise to advance both comprehension and exploration of complex system dynamics, particularly in the intricate realms like optics and photonics.

In this paper, a KAN-based AI4S framework, called as FiberKAN, is proposed for scientific discovery and dynamic characterization of nonlinear fiber optics. Leveraging its powerful nonlinear characterization capability and intrinsic physical interpretability, KANs are employed to unravel the underlying dynamics of physical effects such as power attenuation, chromatic dispersion, and nonlinear effects including self-phase modulation (SPM), self-steepening (SS), and intrapulse Raman scattering (IRS). Through a systematic four-step process of pretraining, pruning, training, and symbolization, KANs are able to accurately obtain explicit, implicit, and non-analytical solutions under various physical effects in fiber-optic systems, demonstrating superior performance compared to MLPs with equivalent parameter scale. The effectiveness, computational cost, and noise resistance of FiberKAN are also evaluated from the perspective of practical applications. Moreover, human intervention such as prior knowledge embedding and hypothesis testing can help better construct KANs for scientific discovery, and KAN's transfer learning ability make it more advantageous than MLP in learning the dynamics of similar systems. Compared with related algorithms including PINN, variant algorithms, and symbolic regression, FiberKAN performs better in accuracy, interpretability, and implementation difficulty. This work demonstrates that KANs are expected to be a promising alternative to MLPs that promotes scientific progress and development of AI4S in nonlinear fiber optics.

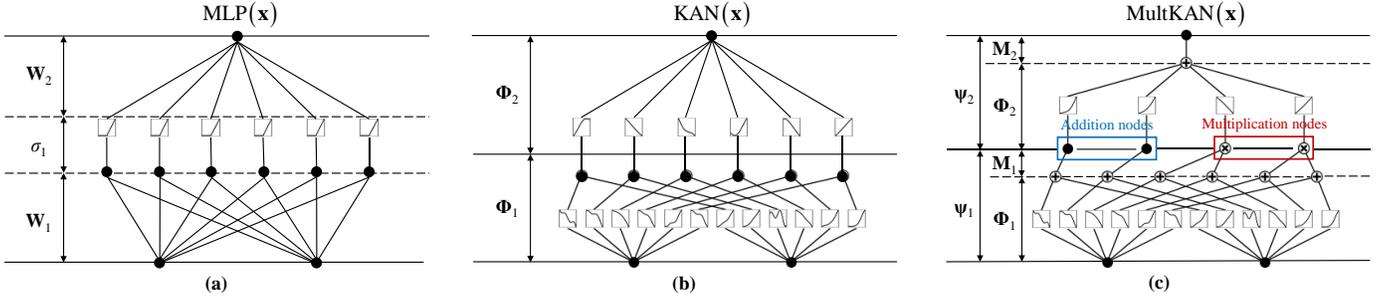

Fig. 1. Comparison between the structures of (a) MLP, (b) KAN, and (c) MultKAN.

## II. METHODOLOGY

### A. Kolmogorov-Arnold Networks

Different from MLPs inspired by the universal approximation theorem [38], the structure of KAN originates from Kolmogorov-Arnold representation theorem [30], which proves that any multivariate continuous function on a bounded domain can be represented as a finite composition of continuous functions of a single variable, and the binary operation of addition. Specifically, if $f(\mathbf{x}) = f(x_1, \ldots, x_n)$ represents a multivariate continuous function defined on a bounded domain, then $f(\mathbf{x})$ can be written as [29]:

$$f\left(\mathbf{x}\right) = f\left(x_1, \cdots, x_n\right) = \sum_{q=1}^{2n+1} \mathbf{\Phi}_q\left(\sum_{p=1}^{n} \phi_{q,p}\left(x_p\right)\right) \quad (1)$$

where $\phi_{q,p}$ is the univariate activation function for variable $x_p$ in the $q^{\text{th}}$ layer, and $\mathbf{\Phi}_q = \{\phi_{q,p}\}$ is a matrix of the activation functions in the $q^{\text{th}}$ layer.

In Eq. (1), it is indicated that learning an arbitrary function essentially boils down to mastering a polynomial number of one-dimensional functions. Based on Kolmogorov-Arnold theorem, it has also been proved that there exist deep networks that can approximate the target function within error $\varepsilon$ ($\varepsilon$ is any scalar in $(0,1)$) for a broad class of continuous multivariate functions [39], which also provides a theoretical proof of the error bound for the subsequent proposal of KAN. Although the activation functions to be learned are finite-dimensional, they may be non-smooth or even fractal, which makes them difficult or even impossible to be learned in practice. Therefore, despite some studies on the original Kolmogorov-Arnold representation theorem [40-41], the extremely complicated function learning process makes it far less well-known than the universal approximation theorem.

Recently, the original Kolmogorov-Arnold representation theorem has been extended to arbitrary depth and width [29,42], and has quickly attracted the interests of scholars in various fields. Unlike the original fixed structure containing two layers with $2n+1$ nodes, the newly proposed KAN structure can flexibly contain any number of layers and nodes, making the one-dimensional function learning process easier and smoother. Accordingly, Eq. (1) can be rewritten as [29]:

$$f\left(\mathbf{x}\right) \approx \sum_{i_{L-1}=1}^{n_{L-1}} \phi_{L-1,i_L,i_{L-1}}\left(\sum_{i_{L-2}=1}^{n_{L-2}} \cdots \left(\sum_{i_1=1}^{n_1} \phi_{1,i_2,i_1}\left(\sum_{i_0=1}^{n_0} \phi_{0,i_1,i_0}\left(x_{i_0}\right)\right)\right)\right) \quad (2)$$

where $L$ denotes the number of layers, $\{n_i\}_{i=0}^{L}$ are the number of nodes (i.e., neurons) in the $i^{\text{th}}$ layer, and $\phi_{i,j,k}$ is the univariate activation function that connects the $k^{\text{th}}$ node in the $i^{\text{th}}$ layer and the $j^{\text{th}}$ node in the $(i+1)^{\text{th}}$ layer. Furthermore, the mapping relationship between the nodes $\mathbf{x}_{i+1}$ in the $(i+1)^{\text{th}}$ layer and nodes $\mathbf{x}_i$ in the $i^{\text{th}}$ layer can be written in the following matrix form:

$$\mathbf{x}_{i+1} = \mathbf{\Phi}_i \mathbf{x}_i = \begin{pmatrix} \phi_{i,1,1} & \phi_{i,1,2} & \cdots & \phi_{i,1,n_i} \\ \phi_{i,2,1} & \phi_{i,2,2} & \cdots & \phi_{i,2,n_i} \\ \cdots & \cdots & \cdots & \cdots \\ \phi_{i,n_{i+1},1} & \phi_{i,n_{i+1},2} & \cdots & \phi_{i,n_{i+1},n_i} \end{pmatrix} \mathbf{x}_i \quad (3)$$

where $\mathbf{\Phi}_i = \{\phi_{i,j,k}\}$ is the matrix of activation functions in the $i^{\text{th}}$ layer. Accordingly, the general form of KAN can be obtained by cascading $\{\mathbf{\Phi}_i\}$, expressed as:

$$\text{KAN}\left(\mathbf{x}\right) = \left(\mathbf{\Phi}_{L-1} \circ \mathbf{\Phi}_{L-2} \circ \cdots \circ \mathbf{\Phi}_1 \circ \mathbf{\Phi}_0\right)\mathbf{x} \quad (4)$$

Intuitively, the structure of KAN in Eq. (4) seems similar to the MLP. To better illustrate the structural differences between MLP and KAN, the comparison between the two structures is displayed in Fig. 1. In MLPs, linear transformations and nonlinear activations are treated separately as $\mathbf{W}$ and $\sigma$, while in KANs, they are combined together as $\mathbf{\Phi}$ for the enhanced interpretability. For this new network paradigm, empirical investigations were conducted to demonstrate that KANs are capable of achieving near-zero training loss in various tasks, and a theoretical analysis and proof were also conducted regarding the convergence of KANs when using gradient descent and stochastic gradient descent [43]. Moreover, the non-zero eigenvalue $\lambda$ of KAN's variant wavelet-based KAN is proved lower bounded by $0.25 \cdot \exp(-b(1-T))^2$, where $b$ and $T$ are the parameters of the activation function applied in the wavelet-based KAN [44]. The decay rate of the eigenvalues can be modulated by adjusting $b$ and $T$ in the activation functions, which provides a means to counter spectral bias. Similarly, the parameters in the activation function in the vanilla KAN can also be adjusted to modulate the decay rate of the eigenvalues, alleviating the spectral bias.

Furthermore, considering the prevalence of multiplication operations in both scientific research and daily life, a multiplicative KAN (MultKAN) integrating the multiplication operation is newly proposed [42], which further improves the interpretability of KANs and potentially simplifies the required network structure. In MultKAN, in addition to the activation function matrix layer $\mathbf{\Phi}$ in the original KAN, the transformation from nodes to those in the subsequent layer incorporates an additional multiplication layer $\mathbf{M}$. The multiplication layer generally comprises two parts: one part handles multiplications between designated multiplication node pairs, while the other applies an identity transformation to addition nodes. If no multiplication nodes are in each layer, the MultKAN will return back to KAN. Similarly, let KAN layer $\boldsymbol{\Psi}=\mathbf{M}\circ\boldsymbol{\Phi}$, the general form of MultKAN can be written as:

$$\text{MultKAN}(\mathbf{x}) = \left(\boldsymbol{\Psi}_{L-1}\circ\boldsymbol{\Psi}_{L-2}\circ\cdots\circ\boldsymbol{\Psi}_1\circ\boldsymbol{\Psi}_0\right)\mathbf{x} \quad (5)$$

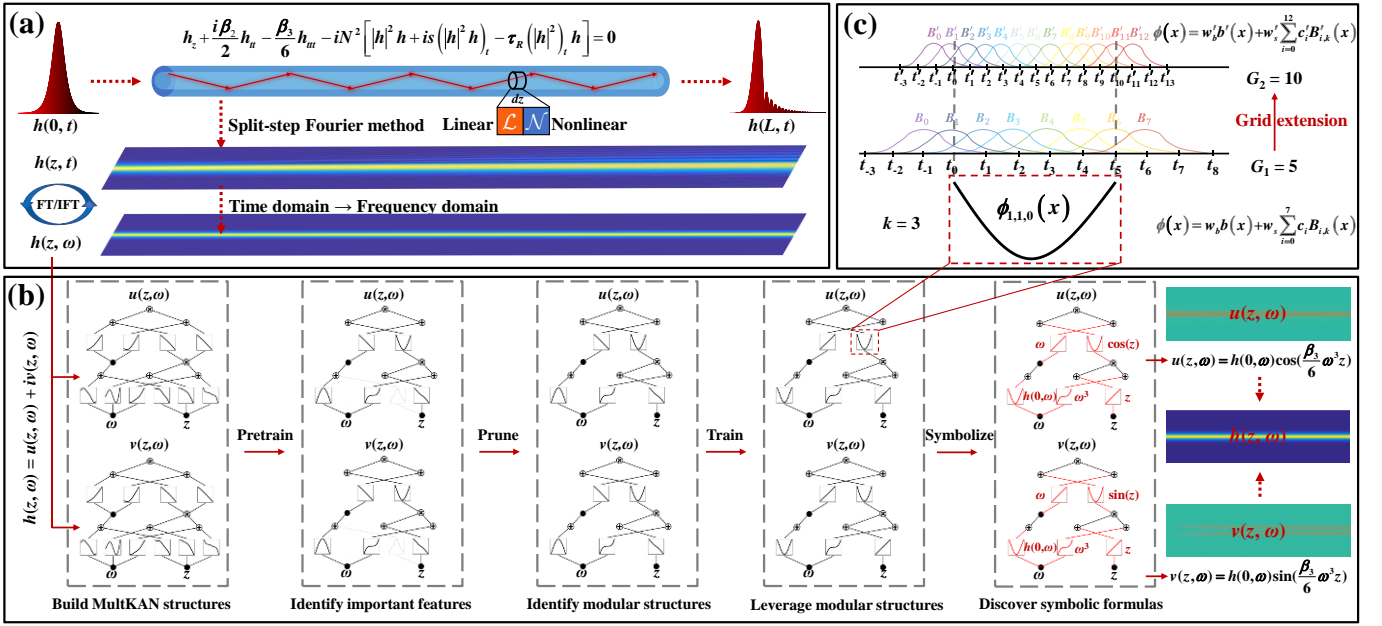

Fig. 2. (a) Nonlinear dynamics in fiber optics and SSFM-based solution. (b) Schematic of KANs for nonlinear dynamics in fiber optics, taking the dynamics learning of pulse propagation under TOD effect as an example. (c) Details of the univariate activation function $\phi_{i,j,k}$, which is parameterized as a B-spline and allows switching between coarse-grained and fine-grained grids.

### B. FiberKAN for nonlinear dynamics in fiber optics

In fiber optics, the propagation of optical pulses in fiber usually suffers from multiple linear and nonlinear physical effects such as group velocity dispersion (GVD), third-order dispersion (TOD), SPM, and higher-order nonlinear effects including SS and IRS, which can be mathematically characterized by the NLSE [11]:

$$h_z + \frac{\alpha}{2}h + \frac{i}{2}\beta_2 h_{tt} - \frac{1}{6}\beta_3 h_{ttt}$$
$$- iN^2\left[|h|^2 h + is\left(|h|^2 h\right)_t - \tau_R\left(|h|^2\right)_t h\right] = 0 \quad (6)$$

where the dimensionless parameters are defined as:

$$h = \frac{H}{\sqrt{P_0}}, z = \frac{Z}{L_D}, t = \frac{T}{T_0}, N^2 = \frac{L_D}{L_{NL}} = \frac{\gamma P_0 T_0^2}{|\beta_2|},$$
$$L_D = \frac{T_0^2}{|\beta_2|}, L_{NL} = \frac{1}{\gamma P_0}, s = \frac{1}{\omega_0 T_0}, \tau_R = \frac{T_R}{T_0} \quad (7)$$

Here, $H(Z,T)$ denotes the complex envelope of a slowly varying optical field, where $Z$ and $T$ are the propagation distance and time scale in a frame of reference moving with the pulse at the group velocity. The dimensionless parameters $h$, $z$, and $t$ are the normalized values of $H$, $Z$, and $T$ under peak power $P_0$, dispersion length $L_D$, and pulse width $T_0$. $N^2$ is the ratio of $L_D$ to the nonlinear length $L_{NL}$, which governs the relative impact of GVD and SPM effects on the pulse evolution along the fiber. The propagation parameters $\alpha$, $\beta_2$, $\beta_3$, and $\gamma$ reflect the power attenuation, GVD, TOD, and the strength of the nonlinear effects. For ultrashort optical pulses with $T_0 < 1$ ps, higher-order nonlinear effects such as self-steepening (SS) and intrapulse Raman scattering (IRS), as expressed by the last two terms of Eq. (6), should be considered, where $s$ and $\tau_R$ are the normalized parameters of SS and IRS, $\omega_0$ and $T_R$ in Eq. (7) are the central angular frequency of the pulse and the first moment of the Raman response function, respectively. Generally, the NLSE in Eq. (6) can be numerically solved with SSFM, as shown in Fig. 2(a). By alternately calculating the nonlinear and linear physical effects in the time domain and frequency domain using inverse Fourier transform (IFT) and FT, the dynamics of pulse propagation in fiber can be approximately characterized.

To achieve accurate, interpretable, and straightforward discovery and characterization of fiber dynamics, two MultKANs are established accordingly as FiberKANs for characterizing the real part $u(z,t)$ / $u(z,\omega)$ and imaginary part $v(z,t)$ / $v(z,\omega)$ of the optical field $h(z,t)$ / optical spectrum $h(z,\omega)$ respectively, where $\omega$ is the frequency scale. Here, we take the dynamics learning of pulse propagation under the TOD effect as an example to illustrate the process of KAN-based fiber dynamics discovery and characterization. Since the propagation laws under TOD effect can be found relatively easily in the frequency domain, two MultKANs are first established to discover the dynamics of $u(z,\omega)$ and $v(z,\omega)$, which takes $z$ and $\omega$ as the inputs, as shown in Fig. 2(b). The activation functions in each layer are initialized to random states. During the pretraining stage, regularization drives the attribution scores of redundant edges toward zero, resulting in a sparse network structure from an initially fully connected network. To better illustrate the optimization process of KANs, we adopt the following strategy to present the results: For edges with extremely small weights $w < 0.01$, they and their connected activation functions will not be shown in the learning details of KANs. For edges with weights $0.01 < w < 0.1$, dotted lines are used to represent their connections with the activation functions, and most of these edges can be removed during the pruning stage. For edges with weights $w > 0.1$, solid lines are used to represent their connections with the activation functions. The transparency of the edges and connected activation functions visually indicates the value of $w$. Moreover, the detailed weight values of edges with $0.01 < w < 0.1$ will be provided in the following results for ease of reading. Moreover, some activation functions become smooth and approximate common basic functions, such as identity, cubic, and trigonometric functions. As the key to implementing KAN-based dynamics learning, the pretraining process plays a pivotal role in identifying both input features and intermediate features. To better capture system dynamics and simplify network optimization, it is natural to prune redundant edges after pretraining. Particularly, when all edges of a node are pruned, the node itself is also removed. The pruning process identifies the modular structure of the physical system, that is, which important features should be connected together. At this point, both the important features and modular structure of the system are determined. Next, further training of the networks are required to improve the accuracy. At this stage, all activation functions become smoother, which allows for an intuitive view of how each node is connected. When the networks are fully trained, symbolization is finally performed. Each activation function will be fitted to the common basic function that is closest to its shape, which allows the mapping relationship between inputs and outputs to be mathematically expressed using symbolic formulas. After symbolization, the networks will further fine-tune the symbolic coefficients, and eventually achieve the machine precision.

During the entire training process, $\Phi$ is crucial for ensuring interpretability and learning high-frequency features. Although multiple basis functions have been applied to construct activation functions $\{\phi_{i,j,k}\}$, we still adopt the B-spline functions in the vanilla KANs to construct the activation function, expressed as:

$$\phi(x) = w_b b(x) + w_s \text{Spline}(x) = w_b \frac{x}{1+e^{-x}} + w_s \sum_i c_i B_{i,k}(x) \quad (8)$$

| Hyperparameter | Value |
|---|---|
| Optimizer | L-BFGS (Part A−Part D) |
| | Adam (Part E) |
| Learning rate lr | 1 (Part A−Part D) |
| | $10^{-3}$ (Part E) |
| $L_1$ regularization coefficient lam_l$_1$ | $10^{-4}$ (Part A−Part C) |
| | 0 (Part D−Part E) |
| Spline order $k$ | 3 |
| Grid intervals $G$ | 5 |
| Base function $b(x)$ | silu |
| Initial injected noise to spline | 0.3 |

where $b(x) = \text{silu}(x)$ and $\text{Spline}(x)$ are the basis function and the Spline function with the trainable scaling parameters $w_b$ and $w_s$, respectively. $B_{i,k}(x)$ denotes the B-Spline function with the trainable weighting parameter $c_i$, and it is governed by the spline order $k$, and the number of grid intervals $G$. To better illustrate the Spline function in the univariate activation function $\phi_{i,j,k}$, a knot vector $\mathbf{T} = \{t_0, t_1, \ldots, t_G\}$ is first defined in Fig. 2(c). Then, the definition of $B_{i,k}(x)$ adopts a recursive method and is divided into the following two cases:

For $k = 1$, $B_{i,k}(x)$ is defined by:

$$B_{i,k}(x) = \begin{cases} 1 & \text{if } t_i \le x < t_{i+1}, i = 1, 2, \ldots, G \\ 0 & \text{otherwise} \end{cases} \quad (9)$$

For $k > 1$, $B_{i,k}(x)$ is defined recursively:

$$B_{i,k}(x) = \frac{x - t_i}{t_{i+k} - t_i} B_{i,k-1}(x) + \frac{t_{i+k} - x}{t_{i+k} - t_{i+1}} B_{i+1,k-1}(x) \quad (10)$$

According to different problems and input activations, the Spline function can be updated from coarse-grained to fine-grained grids when required, as shown in Fig. 2(c).

## III. DEMONSTRATIONS AND RESULTS

To verify the feasibility and performance of FiberKANs in learning nonlinear fiber optics, we systematically follow a route from simple to complex effects, from those with analytical solutions to those without analytical solutions. The datasets of scenarios with analytical solutions (Part A−Part D) are directly generated by calculating their analytical solutions, while the datasets of scenarios with non-analytical solutions are obtained by numerically solving the NLSE using SSFM. The step number of SSFM is set to $100 \, L / L_{NL}$ for the sufficient accuracy. The time and spatial dimension of each dataset are 256 and 201 respectively, and thus the dimension of the solving domain is [256, 201]. For scientific discovery problems (Part A−Part D), only 5000 points randomly selected in the solving domain are applied as the training dataset to discover the analytical solutions. For dynamic characterization problems (Part E), half of the points randomly selected in the solving domain are applied as the training dataset to reconstruct the dynamic process. Moreover, the hyperparameters settings of FiberKANs are listed in Table I, where $L_1$ regularization coefficient lam_l$_1$ = $10^{-4}$ is applied in explicit solution discovery problems to remove redundant edges, while lam_l$_1$ = 0 in implicit solution discovery and nonlinear characterization problems, avoids finding the identity transformation of the solution and better fits the input-output data. When constructing FiberKANs for discovering and characterizing various physical effects in optical fibers, some features of KANs are illustrated simultaneously.

### A. Power Attenuation: Important Feature Screening

As the most fundamental physical effect in fiber optics, power attenuation is ubiquitous in all types of optical fibers. Since power attenuation directly affects the amplitude of the pulse, it manifests itself as a decrease in the amplitude in both time domain and frequency domain. Herein, we study the hyperbolic secant pulse $\text{sech}(t)$ and its attenuation with $z$ in the time domain. Since this process does not involve phase change, only the real part of the pulse needs to be considered, which can be mathematically expressed as $u(z,t) = \exp(-\alpha z) \cdot \mathrm{sech}(t)$ in the time domain, where attenuation coefficient $\alpha = 0.5$ here. As one of the most important effects in optical fibers, the analytical solution of power attenuation describes the effect as a negative exponential decrease in pulse amplitude with distance. The exponential decay term quantifies the amplitude reduction due to fiber attenuation governed by $\alpha$, which reflects material absorption and scattering losses inherent to the fiber. Physically, $\alpha$ is influenced by the fiber's composition (e.g., silica impurities) and operational wavelength. The preservation of the $\mathrm{sech}(t)$ shape confirms that attenuation primarily affects amplitude rather than temporal or spectral broadening, aligning with the linear nature of this effect. To better illustrate the capability of KANs to identify important features and molecular structures, a KAN containing one layer with two multiplication nodes is constructed, as shown in Fig. 3(a). In addition to $z$ and $t$, we also consider adding $\omega$ as a redundant variable in the input layer to analyze KAN's ability to handle dynamics-independent feature.

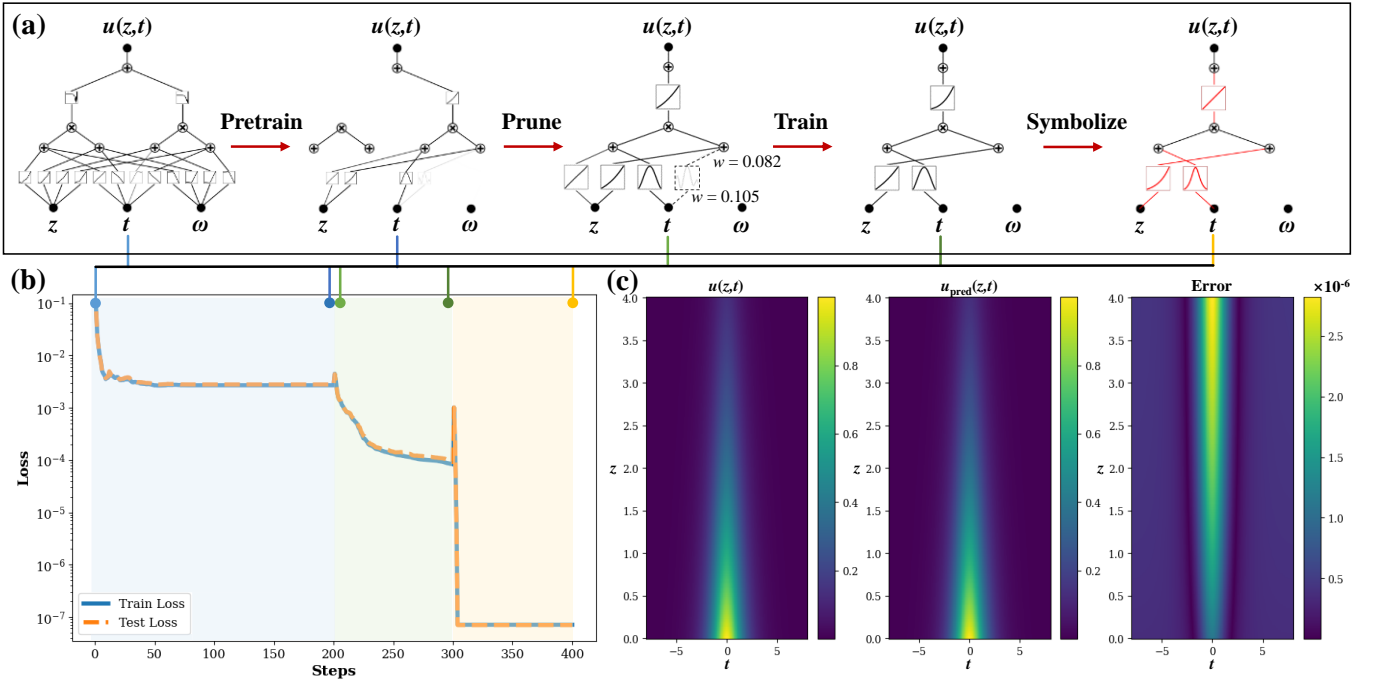

Fig. 3. (a) The learning details of KANs for discovering and characterizing the power attenuation effect. (b) Loss curves of KAN for learning power attenuation. (c) Comparison between the analytical solution and generated solution obtained by KAN, as well as their absolute errors.

As shown in Fig. 3(b), the training loss and test loss exhibit a large decrease at the beginning and gradually converge in the pretraining stage. After pre-training, it can be observed that the node $\omega$ in the input layer has no edges with weights $w > 0.01$ connected to the next layer, indicating that redundant variable $\omega$ is successfully identified as an irrelevant variable in the physical system. Through this mechanism of important feature screening, irrelevant input feature can be intuitively excluded. In addition, a multiplication node in KAN layer $\Psi_1$ has neither input edges nor output edges, which illustrates that it is also redundant. Next, a simpler model with all redundant nodes and edges are pruned is generated for further training. In the stage of further training, the loss curve of the pruned model shows a spike at the beginning, because the pruned nodes and edges also contribute to the reduction of the total loss. As the number of training steps increases, the training loss and test loss further decrease, and eventually reaches the order of $10^{-5}$, which is the maximum accuracy that can be achieved by conventional data-driven DL algorithms. However, the accuracy can be further improved in KANs through symbolic regression. Specifically, the most similar symbol is selected for each edge from the symbol candidate library as the fitting target of the activation function. In this case, the exponential function, identity function and hyperbolic secant function are selected as the fitting target of the remaining three edges. Similarly, the loss curve exhibits a spike after symbolic regression, but soon it reaches the order of $10^{-8}$, which is the machine precision. After rearranging the equation, the equation found by KAN is the same as the actual form, and its coefficients are accurate to the sixth decimal place. As shown in Fig. 3(c), the dynamic evolution generated by KAN is almost completely consistent with analytical solution, with an absolute error of the order of $10^{-6}$, which is far superior to conventional data-driven methods.

### B. Chromatic Dispersion: Deep Structure Discovery

Another typical linear effect in fiber-optic systems is chromatic dispersion, represented by GVD and TOD effects. In SSFM, the chromatic dispersion is calculated in the frequency domain, and when nonlinear terms are not considered, the optical spectrum $h(z,\omega)$ admits an analytical solution. It can be mathematically obtained by solving the ordinary differential equation $ih_z = -i\beta_k\omega^k h / k!$ in the frequency domain, expressed as $h(z,\omega) = h(0,\omega) \cdot \exp(\frac{i}{k!}\beta_k\omega^k z)$, where $k$ is the dispersion order. Given that the FT of the hyperbolic secant function is still a hyperbolic secant function, here we consider $h(0,\omega) = \mathrm{sech}(\omega)$. Accordingly, $u(z,\omega) = \mathrm{sech}(\omega) \cdot \cos(\frac{1}{k!}\beta_k\omega^k z)$, and $v(z,\omega) =$

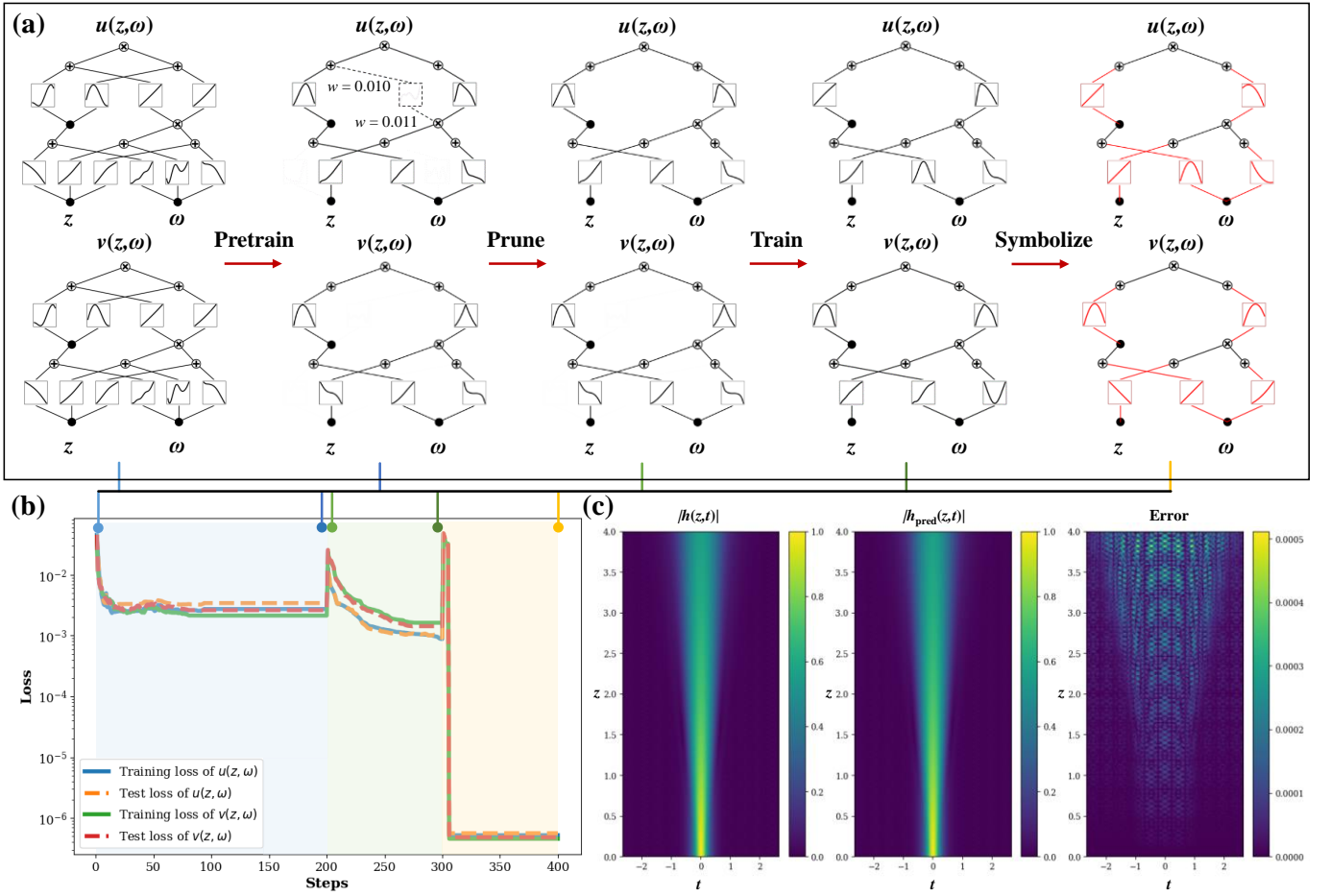

Fig. 4. (a) The learning details of KANs for discovering and characterizing the GVD effect. (b) Loss curves of KAN for learning GVD effect. (c) Comparison between the analytical solution and generated solution obtained by KAN, as well as their absolute errors.

sech$(\omega) \cdot \sin(\frac{1}{k!}\beta_k\omega^k z)$. Only $z$ and $\omega$ are considered as the input variables to simplify the difficulty of learning the underlying dynamics of the system. Unlike the case where only power attenuation is considered, both the real and imaginary parts of the optical field change under the effect of chromatic dispersion. Therefore, two KANs containing one layer with one addition node and one multiplication node are constructed to discover and characterize the dynamics of chromatic dispersion.

When $k = 2$, GVD effect acts alone, the $\beta_2$-dependent term governs pulse broadening due to group velocity mismatch, and the solution show that GVD changes the phase of each spectral component of the pulse by an amount that depends on both the frequency and the propagated distance. Even though such phase changes do not affect the pulse spectrum, they can modify the pulse shape. The details of KANs for learning GVD effect at different training stages is shown in Fig. 4(a). Starting from the same initial structure, both the losses of $u(z,\omega)$ and $v(z,\omega)$ decrease rapidly and gradually converge in the pre-training stage that is similar to the power attenuation, as shown in Fig. 4(b). Since the analytical solutions of $u(z,\omega)$ and $v(z,\omega)$ are the same except the trigonometric operator included, they exhibit the similar modular structures. After 200 steps of pretraining, the KANs for learning $u(z,\omega)$ and $v(z,\omega)$ identify the same features and molecular structure, which is the first step for accurately discovering the dynamics of the real and imaginary parts of the optical spectrum. Through pretraining and pruning, redundant edges are identified and removed. Since the pruned edges have a certain contribution to the original structure, some of the retained edges do not yet exhibit a high similarity to the basic symbols in the symbol candidate library. Subsequently, additional training is performed to further decrease the losses and adjust each activation function. After 100 steps of training, the training and test losses of $u(z,\omega)$ and $v(z,\omega)$ are decreased lower than the coverage values at the pretraining stage. Compared with the pruned structures, the activation functions in the further trained structures are smoother, and the shapes of some activation functions have changed significantly. It is worth noting that some activation functions still have some deviations from the basic symbols, because the limited dataset can be fitted with Splines in various ways. An effective improvement is to add more data to constrain possible activation function combinations. Despite this, KANs is still able to find the most similar symbols for each activation function to fit and successfully find the corresponding analytical solutions in the symbol regression stage. Due to the deviation of activation functions learned in the previous stage, the loss functions in this stage initially show high spikes, but they quickly decrease to the order of $10^{-7}$, indicating that KANs have found the correct dynamics of the system. The comparison between the analytical solution and solution generated by KAN in the time domain is displayed in Fig. 4(c). Unlike the case in power attenuation, the slight differences in the trigonometric coefficients cause the error to show a periodic pattern, but the overall error is still on the order of $10^{-4}$.

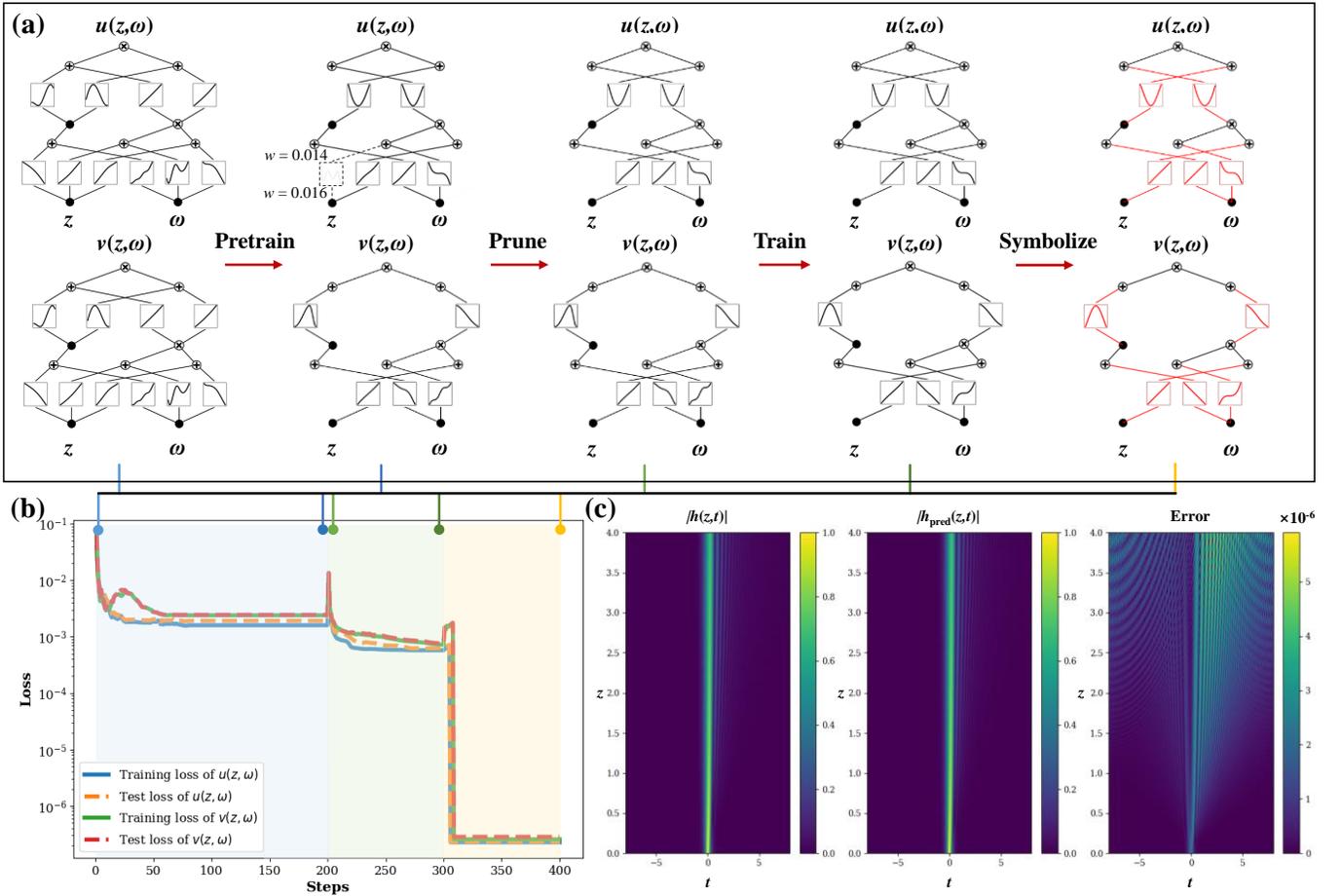

Fig. 5. (a) The learning details of KANs for discovering and characterizing the TOD effect. (b) Loss curves of KAN for learning TOD effect. (c) Comparison between the analytical solution and generated solution obtained by KAN, as well as their absolute errors.

If the pulse wavelength nearly coincides with the zero-dispersion wavelength, higher-order dispersion effects (such as TOD) will provide the dominant contribution to the pulse evolution, leading to pulse skewing and precursor oscillations. The pulse will become asymmetric with an oscillatory structure near one of its edges. To discover the physical laws under the TOD effect, the same KAN structures applied to discover GVD effect are used for TOD dynamics discovery here, and the learning details are shown in Fig. 5(a). Although the analytical solutions of $u(z,\omega)$ and $v(z,\omega)$ have similar forms, it can be observed that the molecular structures identified after pretraining are not exactly the same. Specifically, the positions of the activation functions retained in the output layers of the two structures are completely opposite. In fact, these two structures are equivalent, illustrating that the paths for KANs to learn the system dynamics can be diverse. In Fig. 5(b), the loss curves of $v(z,\omega)$ shows a bump after a rapid decline in the pretraining stage, indicating that the difficulty and process of dynamics learning vary with the learning path. After pruning and further training, the activation functions corresponding to the retained edges become smoother and close to the basic symbols we are familiar with. Naturally, the mathematical expressions of $u(z, \omega)$ and $v(z, \omega)$ can be obtained by symbolic regression. After IFT, the comparison between the solution discovered through KANs and the analytical solution accounting for the TOD effect is displayed in Fig. 5(c). It can be observed that the solution obtained by KAN can accurately characterize the oscillations at the trailing edge of the optical pulse, and the error is mainly concentrated near the oscillations of the optical field. Moreover, the error is on the order of $10^{-6}$, so the TOD effect can also be accurately discovered and characterized.

### C. Optical Soliton: Mixed-effect Dynamics Discovery

In addition to the linear effects, Kerr-based nonlinear effects such as SPM also plays an important role in fiber-optic systems, which leads to the spectrum broadening. When the dispersion length and nonlinear length of the optical pulse are comparable, it is necessary to consider the combined effects of GVD and SPM. New qualitative features arise from the interplay between GVD and SPM in the anomalous-dispersion regime of the fiber-optic system. Specifically, the GVD effect in the anomalous-dispersion regime can suppress the SPM-induced spectrum broadening, and stable or periodic optical solitons are formed under appropriate conditions. Generally, the NLSE has no analytical solutions when GVD and SPM effects coexist, while the analytical solutions of optical solitons can be obtained by the inverse scattering method, and $N$ in Eq. (6) controls the order of soliton. To discover and characterize the dynamics of the optical soliton systems, two KANs are also constructed to learn the real and imaginary parts of the optical fields, respectively.

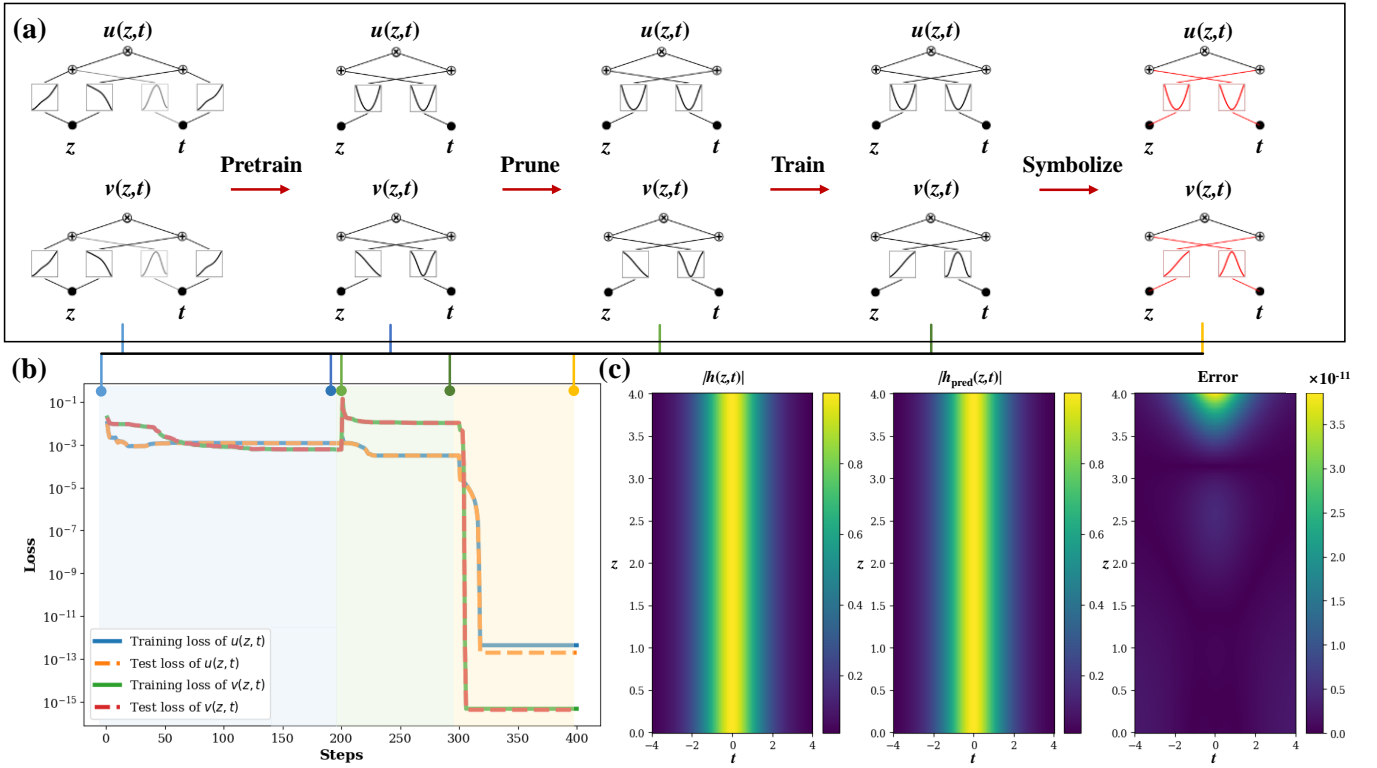

Fig. 6. (a) The learning details of KANs for discovering and characterizing the fundamental soliton. (b) Loss curves of KAN for learning the fundamental soliton. (c) Comparison between the analytical solution and generated solution obtained by KAN, as well as their absolute errors.

When $N = 1$, the GVD and SPM effects suppress each other and exhibit a stable and unchanged evolution in the optical fiber. Such fundamental soliton has a concise analytical solution form, expressed as $h(z,t) = \text{sech}(t) \cdot \exp(iz/2)$. The phase term $\exp(iz/2)$ represents the nonlinear phase accumulation of the fundamental soliton dynamics, which indicates that the phase of the fundamental soliton changes periodically with a period of $z = 4\pi L_D$. Unlike building deeper and wider structures to enhance learning capabilities when adopting neural networks, the construction of KANs often starts from simple structures, which is the mindset of physicists to try the simplest thing first. Generally, the dynamics in nature are presented in the form of relatively simple equations and solutions, and simple structures could help us discover them more efficiently. Moreover, simple structures reduce overfitting risks and align with Occam's razor (i.e., entities should not be multiplied unnecessarily) in knowledge discovery [45], ensuring models capture only necessary dynamics. Moreover, physical prior knowledge and human intuition of dynamic evolution can also help us make a preliminary determine a more appropriate KAN structure. From the evolution of the fundamental soliton, it can be observed that the pulse amplitude remains unchanged, and the phase changes periodically. Our prior knowledge and empirical intuition guide us that there will be a periodic term in the analytical solution of real and imaginary parts that changes periodically with the propagation length $z$, and the overall molecular structure of the solution will not be too complicated. Accordingly, simple KAN structures are constructed, as shown in Fig. 6(a). Since the initially established KANs have few redundant edges, the loss curves of $u(z,t)$ and $v(z,t)$ in the pretraining stage converge quickly, as shown in Fig. 6(b). It can be seen that the weights of the redundant edges are almost 0, and the initial protrusion is relatively small when training the pruned structures. It is worth noting that the activation functions of KAN for $v(z,t)$ show the opposite sign after training, but the loss functions are reduced to a lower level. After symbolization, the losses of both $u(z,t)$ and $v(z,t)$ are reduced to below the order of $10^{-10}$, indicating that simpler structures can more keenly capture the system dynamics. The comparison between the analytical solution and generated solution are shown in Fig. 6(c), where the absolute error is only on the order of $10^{-11}$.

### D. Self-steepening Effect: Implicit Dynamics Discovery

When the pulse width is less than 1.0 ps, higher-order nonlinear effects are required to be considered. As a typical higher-order nonlinear effect, SS effect results from the intensity dependence of the group velocity, leading to an asymmetry in the SPM-broadened spectra of ultrashort pulses. Generally, the evolution of ultrashort pulse under SS effect cannot be analytically solved. However, it has an intensity-phase separated analytical solution when the attenuation and dispersion are neglected. The general solution of this special case is $I(z,t) = f(t-3sIz)$, where $I$ is the intensity of $h(z,t)$, and $f(\cdot)$ describes the pulse shape at $z = 0$. The solution under SS effect illustrates that each $t$ moves along a straight line from its initial value, and the slope of the line is intensity dependent, which leads to pulse distortion. Physically, the group velocity of the pulse is intensity dependent such that the peak moves at a lower speed than the wings. As a result, the trailing edge becomes steeper and steeper with increasing $z$. Unlike the previous cases, it is an implicit solution without explicit function expression. For most scientific discovery methods, implicit physical laws are usually difficult to identify. It is expected to discover such implicit solution based on the special structure of KANs. Here, we still set $h(0,t) = \mathrm{sech}(t)$, and $f(\cdot) = \mathrm{sech}(\cdot)$ accordingly. To discover the implicit law, $I$ is also applied as an input feature, and the regularization term is removed to prevent the network from learning the direct mapping of $I = I$.

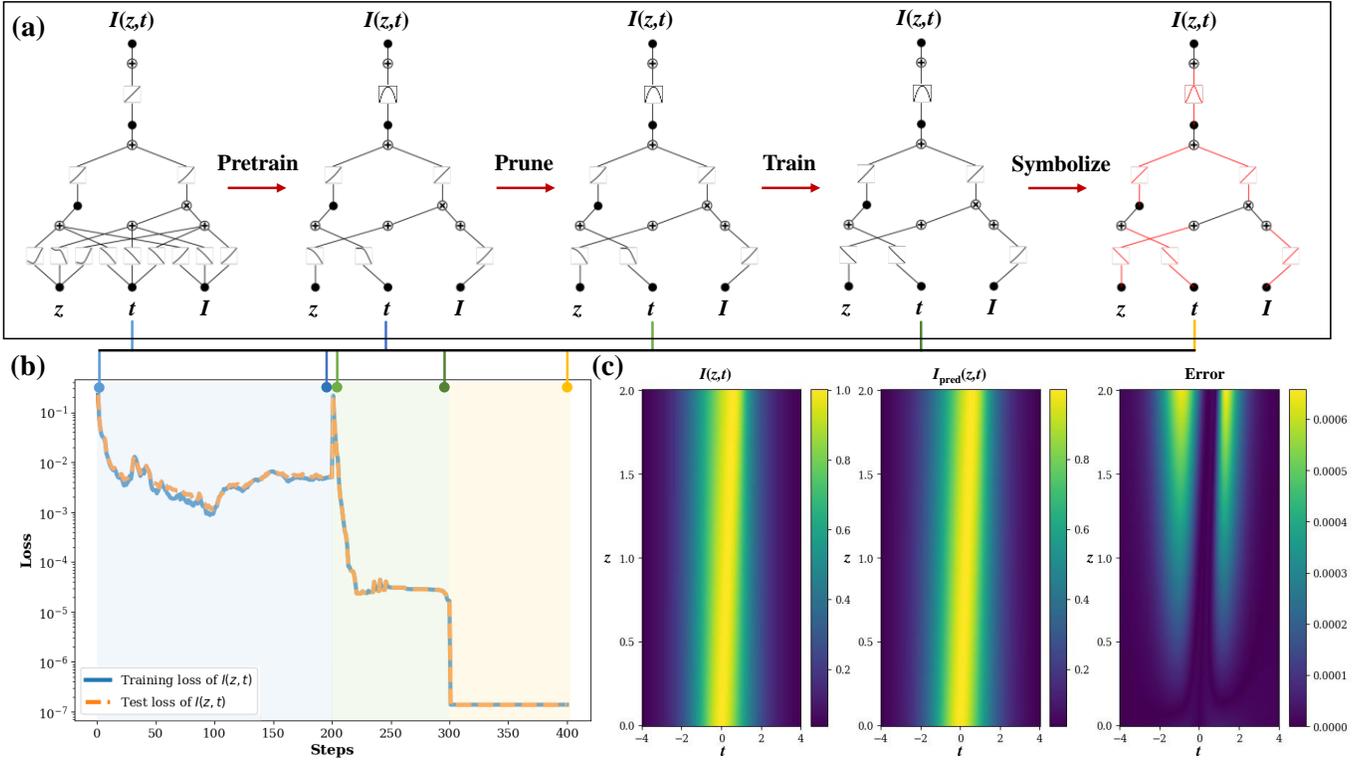

Fig. 7. (a) The learning details of KANs for discovering and characterizing the SS effect. (b) Loss curves of KAN for learning SS effect. (c) Comparison between the analytical solution and generated solution obtained by KAN, as well as their absolute errors.

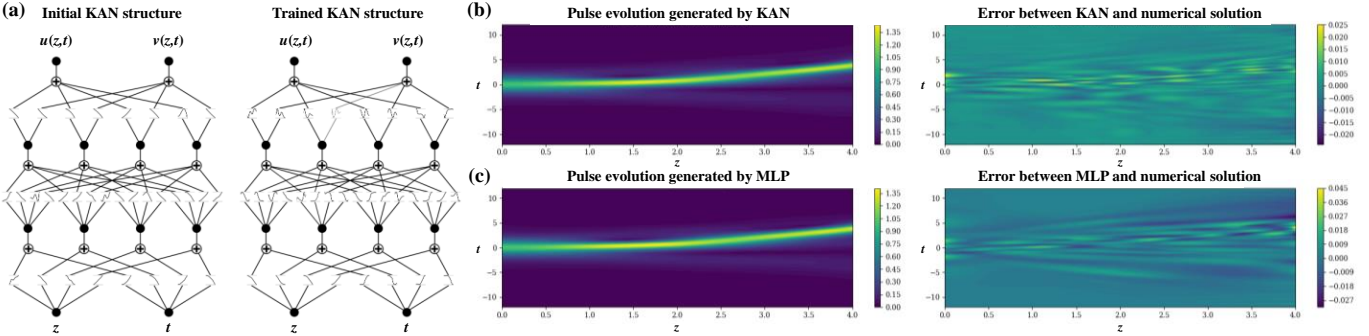

Fig. 8. (a) The initial and trained KAN structures for characterizing the IRS effect. (b) Pulse evolution generated by KAN and the corresponding error distribution relative to the numerical solution. (c) Pulse evolution generated by MLP and the corresponding error distribution relative to the numerical solution.

A KAN with three input features is constructed to learning SS effect, and its optimization details are illustrated in Fig. 7(a).

Although the network structure is not complicated, a large number of redundant edges between the input layer and the first KAN layer need to be removed. Due to the absence of the regularization term, the training loss and test loss of the network are more volatile in the first 200 steps, as shown in Fig. 7(b). Nevertheless, the established KAN still correctly finds the required edges. After pruning, KAN finds the correct activation function for each edge with several training steps, and the loss drops to a fairly small value. Since each activation function has been optimized to approximate the corresponding basic symbol, the symbolization process does not experience a large mutation at the beginning, which is similar to the case of optical soliton. As shown in Fig. 7(c), the pulse becomes asymmetric with its peak shifting, and its trailing edge becomes steeper and steeper with the increasing $z$. In this implicit solution case, the complicated mapping relationship between inputs and output

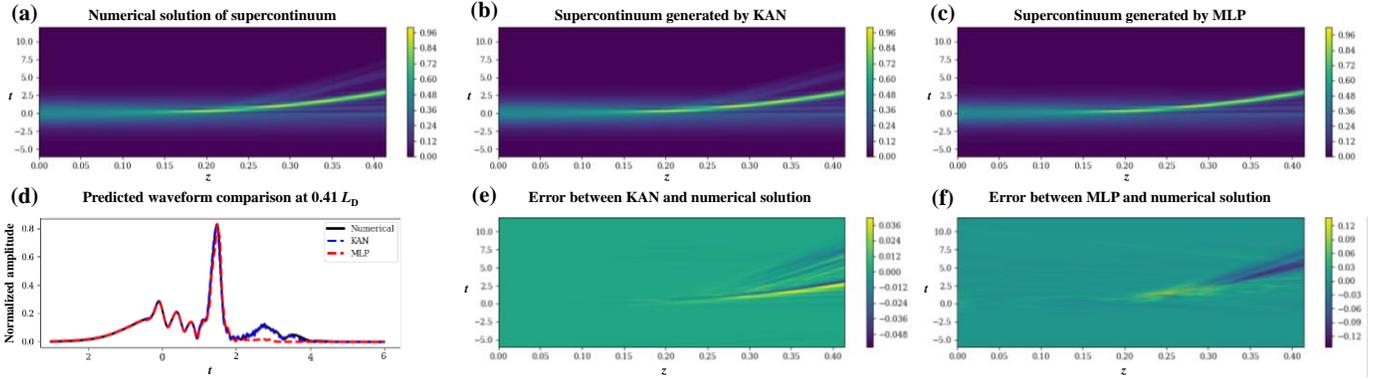

Fig. 9. (a) Numerical solution of supercontinuum. Supercontinuum generated by (b) KAN and (c) MLP. (d) Predicted waveform comparison at $0.41L_D$. (e) Error between KAN and numerical solution. (f) Error between MLP and numerical solution.

can still be accurately identified and characterized by KAN.

### E. Intrapulse Raman Scattering: Non-analytical Dynamics Characterization

In addition to SS, IRS is another typical higher-order nonlinear effect in ultrashort pulse evolution, which will lead to the soliton self-frequency shift. Physically, IRS appears as a red shift increasing with distance of the soliton spectrum. Since the resulting perturbation of the soliton is non-Hamiltonian type when IRS is considered, it lacks an analytical pulse-like solution [46]. In this case, it is unable to symbolically characterize the pulse evolution as in the previous cases. Nevertheless, it is still possible to construct KANs to learn the dynamics of IRS effect in a data-driven manner similar to MLPs.

Herein, a KAN consisting of two layers with four nodes in each layer is constructed to characterize the dynamics of IRS effect, as shown in Fig. 8(a). For comparison, an MLP containing two hidden layers with 22 neurons in each layer is also constructed for the IRS dynamics learning. Since the sigmoid activation function ($\sigma(x) = 1 / (1 + e^{-x})$) is structurally similar to the silu function applied in the basis function $b(x)$ in Eq. (8), here the sigmoid function is applied as the activation function of MLP. In this configuration, the two structures have a similar number of trainable parameters (608 in KAN model, and 618 in MLP model), making them have almost the same network scale. After 10000 iterations, the trained KAN structure is displayed in Fig. 8(a). Although each activation function in KAN does not approach a certain basic symbol as in the previous cases, the trained KAN can establish a complex mapping from $(z,t)$ to $(u,v)$ through a series of addition operations, which follows the Kolmogorov-Arnold representation theorem. As shown in Fig. 8(b), the evolution generated by KAN accurately exhibit the dynamic process of energy transfer from blue-shifted components to red-shifted components, and the maximum error is 0.0241. As a comparison, the pulse evolution results and error distribution generated by MLP after 10,000 iterations are shown in Fig. 8(c). Although the constructed MLP contains more trainable parameters than KAN, it presents a larger learning error under the same training conditions. This indicates that KAN performs superior learning and representation capabilities at the comparable trainable parameter scale. Therefore, as a new network paradigm, KAN is not only capable of scientific discovery of potential analytical solutions, but also has great potential in data-driven dynamic characterization tasks.

Furthermore, for ultrashort pulses propagation in ultrafast optics, they usually experience a mixture of all the above-mentioned linear and nonlinear physical effects, leading to the phenomena of supercontinuum generation [47]. In addition to the soliton splitting, higher-order dispersion and nonlinearity will induce high-frequency dynamics in supercontinuum, which is difficult to be learned by traditional MLPs [48]. Here we study the dynamics learning of supercontinuum under the combined action of all the above-mentioned effects to evaluate KAN's ability in high-frequency nonlinear dynamics learning. Given that the ultrafast optical system is more complex, a larger KAN is established for dynamic characterization, which contains three KAN layers with 10 addition nodes per layer (3360 trainable parameters in total). Accordingly, an MLP with three hidden layers and 40 neurons per layer is also established, which has 3482 trainable parameters. In this scenario, the supercontinuum generation of a hyperbolic secant pulse with $P_0$ = 1500 W and propagation length of 0.41 $L_D$ is considered. After training, the comparison between the generation results of KAN and MLP is shown in Fig. 9. Compared with the numerical results in Fig. 9(a), supercontinuum generated by KAN in Fig. 9(b) exhibits better similarity than supercontinuum generated by MLP in Fig. 9(c), especially in the high-frequency components. The learning ability of KAN and MLP for different frequency components can be more intuitively observed from the waveform comparison in Fig. 9(d), where both methods learn well on the low-frequency components, while MLP can hardly learn the high-frequency components. Furthermore, the maximum error of KAN in the entire domain is almost one order of magnitude higher than that of MLP, indicating that KAN can better learn the dynamics of complex systems with the equivalent parameter scale, as shown in Fig. 9(e) and Fig. 9(f).

## IV. COMPREHENSIVE EVALUATION OF FIBERKAN IN PRACTICAL APPLICATIONS

Different from previous simulation tests, the practical application of FiberKAN needs to consider several aspects of performance. In this section, a comprehensive evaluation of FiberKAN is conducted from the perspective of practical application. Specifically, we will explore the potential of FiberKAN in practical applications from aspects such as effectiveness, computational cost, and noise resistance.

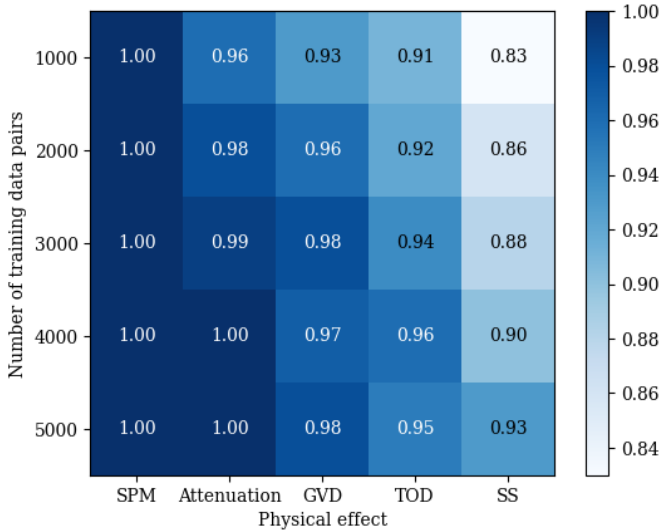

Fig. 10. The success rate of KAN-based scientific discovery method in discovering dynamics under different physical effects with different numbers of training data pairs.

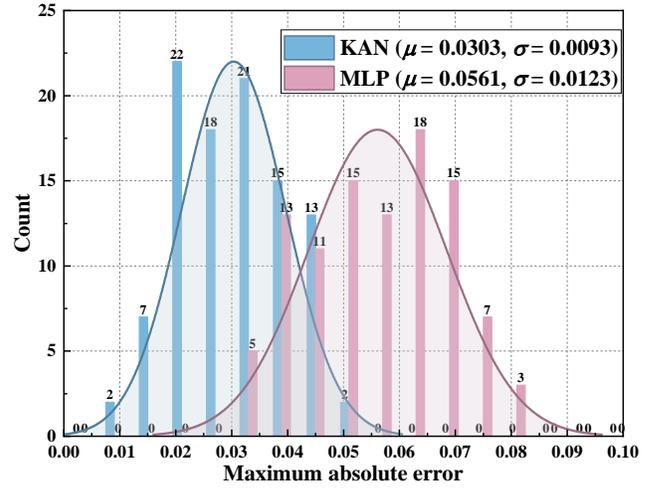

Fig. 11. Comparison of statistical tests between KAN and MLP in characterizing the IRS effect.

### A. Effectiveness of FiberKAN

In practical applications, an important issue is the success rate of KANs in discovering the exact symbolic formula and statistical error of KANs in characterizing the high-order nonlinear effects, which determines the effectiveness of FiberKAN in practical nonlinear fiber optics.

Different from selecting correct terms from a candidate library in conventional scientific discovery methods, trainable activation functions allow the network to learn the system dynamics in an arbitrary way, leading to some local optimal cases where some activation functions cannot be symbolized. To verify the effectiveness of KANs in discovering the dynamics of fiber-optic systems, statistical tests are conducted on the dynamics discovery under different physical effects. Simultaneously, different numbers of training data pairs are set to analyze the performance of KANs with different data amounts. For each case, 100 independent tests are conducted, and the statistical results are shown in Fig. 10. It can be observed that the increase in the amount of training data will help accurately identify the molecular structure of the dynamics in most cases, although it will increase the training time of KANs. Due to the simplicity of the analytical solution for the soliton, the success rate of identifying its dynamics remains at 100% in different cases. In addition, the success rate remains above 95% in the scenario where power attenuation exists. However, the success rate of dynamics learning cannot remain 100% under the effect of GVD or TOD, which illustrates that KAN may fall into local optimal solutions when learning complex dynamics, even with a large amount of data. Moreover, the performance of KANs declines in learning dynamics under the SS effect. In most failed cases, KANs discover the direct mapping relationship of $I = I$, indicating that the dynamics of implicit solution is more difficult to identify when the amount of data is small. Nevertheless, KANs can still achieve a success rate of 93% when the number of data pairs is 5000. Statistical results show that KANs are able to accurately discover fiber dynamics without any prior knowledge or human intervention in most of time. In fact, some of our prior knowledge and empirical intuition about the system can further improve the effectiveness of scientific discovery by guiding the design and optimization of KAN's structure and training, which will be detailed in Part A and Part B of Section V.

Moreover, although the results of Fig. 8 and Fig. 9 show that the KAN performs better than MLP in nonlinear representation problems, the statistical performance of KAN has not been tested. Without loss of generality, here we conduct statistical tests on the scenario in Fig. 8. Specifically, the same KAN structure and MLP structure in part E of Section III are used to ensure that they are trained with similar scale of trainable parameters. The two models are trained to convergence with 100 different random seeds, and maximum absolute errors of these models in this scenario are counted. The statistical test results in characterizing the IRS effect are shown in Fig. 11. The mean value $\mu$ of the maximum absolute error obtained by MLP is 0.0561, and the standard deviation $\sigma$ is 0.0123. While $\mu$ = 0.0303 and $\sigma$ = 0.0093 using KAN. Although the error range of KAN overlaps with that of MLP, the results of KAN is better in most cases, and the lower standard deviation also illustrates that KAN is more stable in nonlinear characterization problems. In addition, when the nonlinearity is stronger and the frequency is higher, the advantage of KAN using trainable activation functions will be more obvious, which can be seen from Fig. 9.

### B. Computational Cost of FiberKAN

Another important issue of KANs in practical applications is the computational cost. The different treatments of activation functions in KAN and MLP structures make their structures present different characterization capabilities and complexities. The two structures with similar scales of trainable parameters are compared in Part E of Section III, which is actually a comparison from the perspective of memory cost, and KAN shows stronger characterization capabilities under the similar memory usage. However, for practical deployment in large-scale fiber systems, the cost of running time is more concerned. Under the same hardware device, the running time of the algorithm is mainly determined by the number of operations included in the algorithm. Therefore, we will compare the running time of the two structures according to the number of operations.

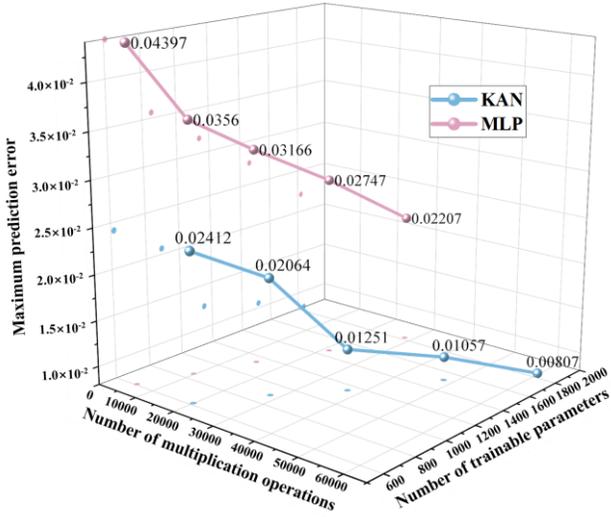

Fig. 12. Maximum prediction error of KANs and MLPs with different number of multiplication operations and similar scale of trainable parameters.

TABLE II
AVERAGE RUNNING TIME OF KAN AND MLP

| Structure | Number of multiplication operations | Number of trainable parameters | Average training time of one iteration (s) | Inference time (s) |
|---|---|---|---|---|
| KAN | 19936 | 608 | 0.212 | 0.0035 |
| | 28035 | 855 | 0.236 | 0.0040 |
| | 37380 | 1140 | 0.244 | 0.0046 |
| | 47971 | 1463 | 0.258 | 0.0051 |
| | 59808 | 1824 | 0.271 | 0.0055 |
| MLP | 5324 | 618 | 0.0178 | 0.0017 |
| | 9660 | 893 | 0.0185 | 0.0018 |
| | 14504 | 1149 | 0.0194 | 0.0019 |
| | 21240 | 1514 | 0.0205 | 0.0021 |
| | 28184 | 1842 | 0.0212 | 0.0038 |

For a general MLP structure with $L$ hidden layers, its mapping relationship from inputs to outputs can be represented by:

$$\mathbf{H}_1 = \sigma\left(\mathbf{W}_1\mathbf{X} + \boldsymbol{b}_1\right),$$
$$\mathbf{H}_j = \sigma\left(\mathbf{W}_j\mathbf{H}_{j-1} + \boldsymbol{b}_j\right), \text{ for } j = 1, 2, ..., L. \quad (11)$$
$$\mathbf{O} = \mathbf{W}_{L+1}\mathbf{H}_L + \boldsymbol{b}_{L+1}.$$

where $\mathbf{X}$, $\mathbf{H}_j$, and $\mathbf{O}$ are the input layer, hidden layer, and output layer feature vector, and $j$ denotes the index of the hidden layer. $\mathbf{W}_j$, $b_j$, and $\sigma$ are the weight matrix, bias vector, and activation function, respectively. For sigmoid activation function, it contains three operations (1 exponential, 1 addition, and 1 division operation). Generally, the running time of division operation is approximately equal to that of multiplication, and exponential operation are equivalent to at least 5 multiplication operations to ensure sufficient accuracy when using Taylor expansion and ignoring addition operations. If all hidden layers have the same number of neurons, the MLP structure includes at least $10n_xn_w + 10\left(L-1\right)n_w^2 + n_wn_o$ multiplication operations and $2Ln_w + n_o$ addition operations, where $n_x$, $n_w$, $n_o$ are the number of neurons in the input, hidden, and output layers.

For a general KAN structure, its mapping relationship from inputs to outputs can be represented by Eq. (2) − Eq. (4). In this structure, the B-Spline function $B_i(x)$ in the activation function $\phi(x)$ shown in Eq. (8) of the original manuscript contributes the most operations. According to the recursive expression shown in Eq. (10), $B_{i,k}(x)$ contains $5(k-1)$ addition operations and $4(k-1)$ multiplication operations. Therefore, $\text{Spline}(x)$ in Eq. (8) contains $(G+k-1)\cdot 4(k-1)\cdot(G+2k)$ multiplication operations (It requires $(G+2k)$ $B_{i,k-1}(x)$ to calculate all $B_{i,k}(x)$, as shown in Fig. 2(c).), and $(G+k)\cdot 5(k-1)\cdot(G+2k)$ addition operations. Considering the basis function $b(x)$ in Eq. (8), one activation function $\phi(\mathrm{x})$ contains at least $4(G+k-1)\cdot(k-1)\cdot(G+2k)+7$ multiplication operations, and $5(G+k)\cdot(k-1)\cdot(G+2k)+1$ addition operations. If all nodes in KAN are addition nodes, the numbers of multiplication and addition operations should be multiplied by $n_xn_w + (L-1)n_w^2 + n_wn_o$.

More specifically, we calculate the number of operations in KAN and MLP structures used in Fig. 8. For MLP, $L = n_x = n_o = 2$, $n_w = 22$, and it requires at least 5324 multiplication operations and 90 addition operations. For KAN, $L = n_x = n_o = 2$, $n_w = 4$, $G = 5$, $k = 3$, and it requires at least 19936 multiplication operations and 881 addition operations. Since the complexity of addition operations is much smaller than that of multiplication operations in modern hardware devices, the running time is mainly determined by the number of multiplication operations. Therefore, the time cost of KAN is about 3.74 times that of MLP in this case. Next, we further test the two-layer KANs with 4−8 nodes per layer and MLPs with equivalent scale of trainable parameters on this case, and calculate the number of multiplication operations for comparison. In Fig. 12, it can be observed from the first two points of the KAN curve and the last two points of the MLP curve that when the number of multiplication operations is similar, the maximum prediction error of KAN is still smaller than that of MLP, which also illustrates the powerful characterization ability of KAN.

In fact, the multiplication operations mainly measure the complexity of the forward propagation of two structures, which also measures the inference complexity. However, the training stage includes not only forward propagation but also backward propagation for updating trainable parameters, which requires further calculating the gradient of the loss function with respect to each trainable parameter. For MLP, the trainable parameters include the weight matrix $\mathbf{W}_j$ and bias vector $b_j$ of each hidden layer, while the trainable parameters in KAN are the scaling parameters $w_b$ and $w_s$, as well as the weighting parameter $c_i$ in the activation function $\phi(x)$ of each node. In the backward propagation of MLP, the classic chain rule is usually used to calculate the gradient, which also applies to the KAN structure. However, the multiplication operations involved in backward propagation chain of KAN are 2−5 times more than those in MLP, and the current optimization algorithms will further improve the training speed of the MLP structure. Therefore, when the scale of trainable parameters is the same, KAN is indeed more complex to train than MLP. To better illustrate the computational overhead of the two structures, we calculate the average running time of the two methods applied in Fig. 12, as shown in Table II. It can be seen that the inference time of KAN and MLP is almost the same, while the training time is slower than MLP due to more operations and optimization algorithms in the backward propagation. Nevertheless, the computational overhead of KAN is acceptable, especially when using the trained model for inference. Moreover, if the interpretability of KAN is not the focus, multiple variants of KAN that have been proposed can reduce the running time, such as the efficient KAN [32], Kolmogorov-Arnold Fourier Networks [33], and ReLU-KAN [34], which using Chebyshev polynomials, random Fourier features, and ReLU function to construct activation functions in KAN, respectively.

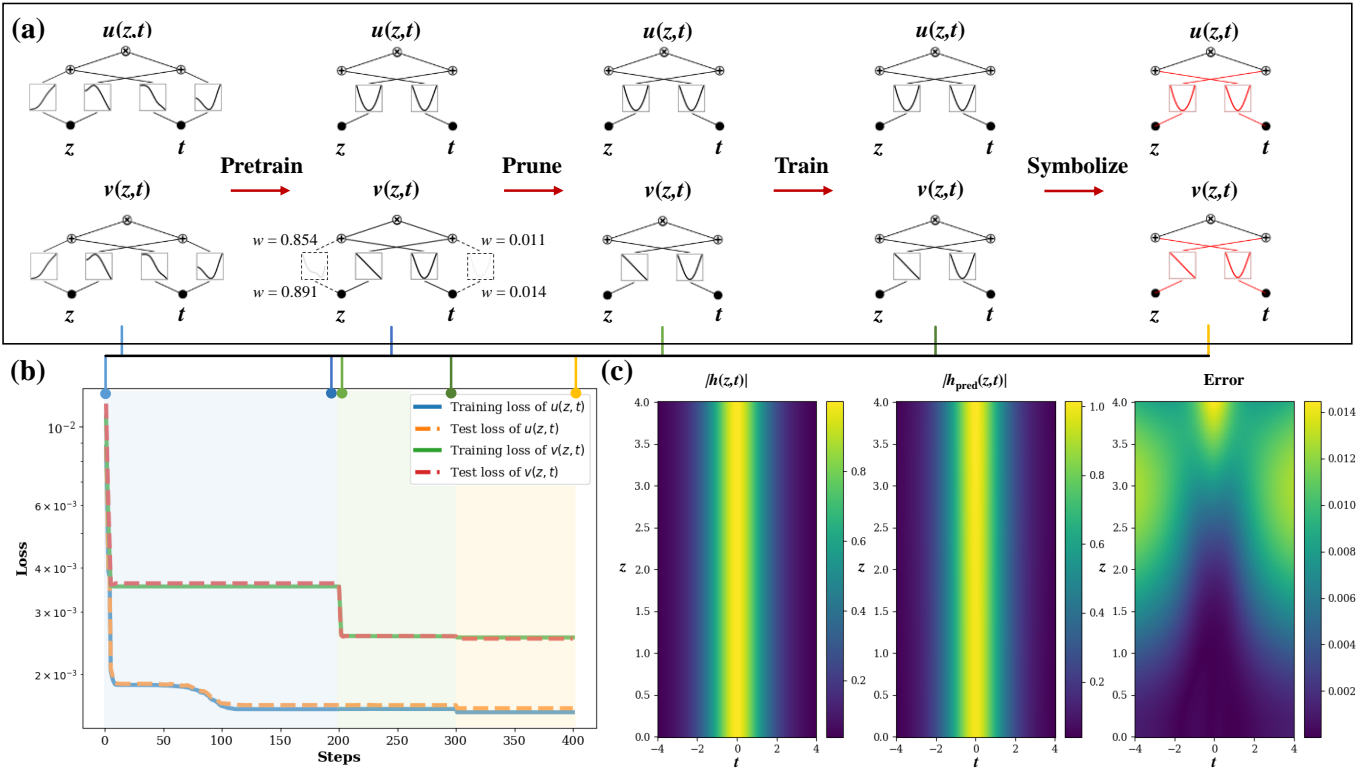

Fig. 13. (a) The learning details of KANs for discovering and characterizing the fundamental soliton with 1% Gaussian noise. (b) Loss curves of KAN for learning the fundamental soliton with 1% Gaussian noise. (c) Comparison between the analytical solution and generated solution obtained by KAN, as well as their absolute errors.

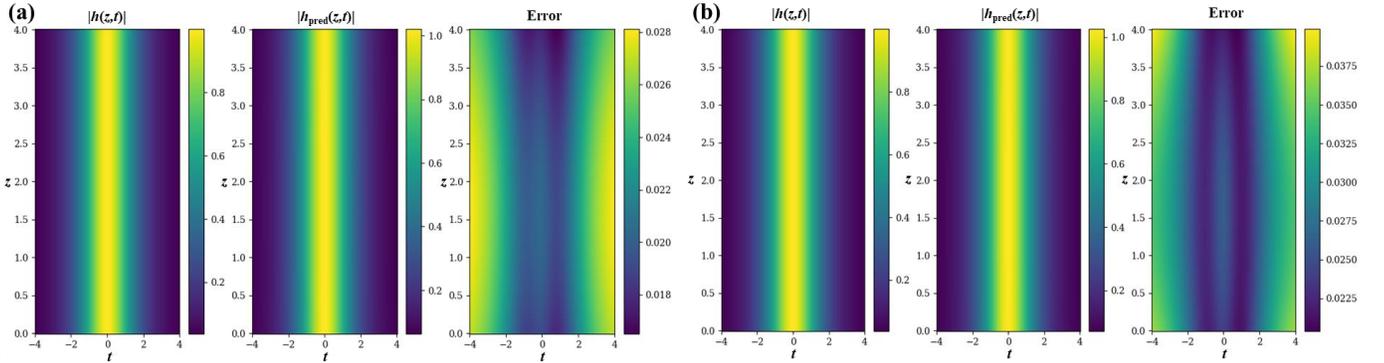

Fig. 14. Comparison between the analytical solution and generated solution obtained by KAN, as well as their absolute errors, using the fundamental soliton data with (a) 2% Gaussian noise, and (b) 3% Gaussian noise.

### C. Noise Resistance of FiberKAN

In previous verifications of the scientific discovery problems, noise is not considered in the data. However, the noise will directly affect the accuracy of the analytical solution discovered. To explore the ability of FiberKAN to discover the dynamics in noisy data, here we add different degrees of noise to the analytical solution of the fundamental soliton that applied as the training data. First, considering adding 1% Gaussian noise to the analytical solution of the fundamental soliton, and the same KAN structure and training approach in Part C of Section III are applied to learn the dynamics in the noisy data. As shown in Fig. 13(a), the learning process of KAN for the real part $u(z,t)$ of the fundamental soliton is almost the same as that in Fig. 6 (a), but the learning of the imaginary part $v(z,t)$ exhibits a different path. Nevertheless, the remaining two activation functions are still correctly identified as the corresponding functions in the symbolization stage. Since the solution discovered by FiberKAN cannot provide an analytical solution to the additional noise, the training and test losses in the symbolization stage of Fig. 13(b) cannot be reduced to the machine precision. Correspondingly, the error of solution generated compared with the analytical solution cannot reach machine precision. Despite the decrease in accuracy, KAN still accurately discovered the analytical solution of the fundamental soliton from the noisy data, which reflects the good noise resistance of FiberKAN. Furthermore, we consider adding higher noise to the analytical solution of the fundamental soliton, and the generated results using data with 2% and 3% Gaussian noise are shown in Fig. 14(a) and Fig. 14(b), respectively. Similarly, the soliton solutions under both noise conditions can be learned accurately. In particular, if the analytical solution is added with the corresponding 2% and 3% noise, the error of the solutions generated by FiberKAN are only on the order of $10^{-3}$, which also illustrates its good noise resistance.

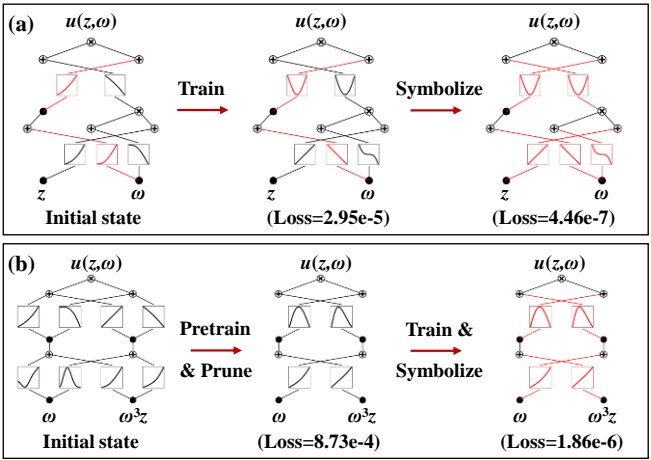

Fig. 15. Prior knowledge embedding of KANs for learning $u(z,\omega)$ under the TOD effect: (a) Molecular structure presetting; (b) Auxiliary variables presetting.

## V. Discussion

As a freshly proposed network with distinct structure, KAN is constructed and trained fundamentally different from MLP, and many intuitions and insights for MLPs may not be directly applicable to KAN. Consequently, its distinctive attributes bring both challenges and opportunities for scientific discovery and dynamic characterization within the domain of nonlinear fiber optics. In this section, human intervention approaches such as prior knowledge embedding and hypothesis testing are investigated to improve the effectiveness of KANs in scientific discovery. In addition, the transfer learning of KANs are discussed to illustrate their advantages in learning the dynamics of similar effects in fiber-optic systems. Moreover, we also discuss the characteristics of KAN compared with some related algorithms.

### A. Prior Knowledge Embedding of KAN

In addition to the trainable activation functions, an important feature that distinguishes the training of KAN from MLP is that its training process allows more human intervention by presetting partially known molecular structures or auxiliary variables. This interactivity allows prior knowledge to be embedded, which helps to make dynamic characterization more accurate and faster. Since some complex structures or features are artificially embedded to the training process, the system-level problem is simplified to a subsystem-level problem, thereby reducing the learning difficulty. To better illustrate the prior knowledge embedding of KAN, dynamics learning of $u(z,\omega)$ under the TOD effect is selected as an example, because its learning process including higher-order coupling term $\omega^3 z$. Specifically, molecular structure and auxiliary variables presetting are performed for learning $u(z,\omega)$ under the TOD effect, where the prior knowledge of input pulse shape and coupling term are respectively considered, as shown in Fig. 15(a) and Fig. 15(b). When learning the dynamics of hyperbolic secant pulses, it is natural to consider $\mathrm{sech}(t)$ or $\mathrm{sech}(\omega)$ as part of the solution according to the prior knowledge, so the molecular structure containing sech($\omega$) can be preset as part of KAN structure, as shown in Fig. 15(a). After training from the initial state, the total loss in the training stage is an order of magnitude lower than that in Part B of Section III. Except for the fixed identity and hyperbolic secant functions, other activation functions are closer to the corresponding basic functions compared with that in Part B of Section III, and lower loss value is further achieved in the symbolization stage. In addition, the combinations of independent variables can be set as auxiliary variables to reduce the difficulty of KANs in discovering complex molecular structures. Here, $\omega^3 z$ is set as an auxiliary variable for TOD effect discovery, as shown in Fig. 15(b). By setting auxiliary variables, KAN is able to discover the dynamics of TOD effect with simpler molecular structures, thereby increasing the success rate of scientific discovery.

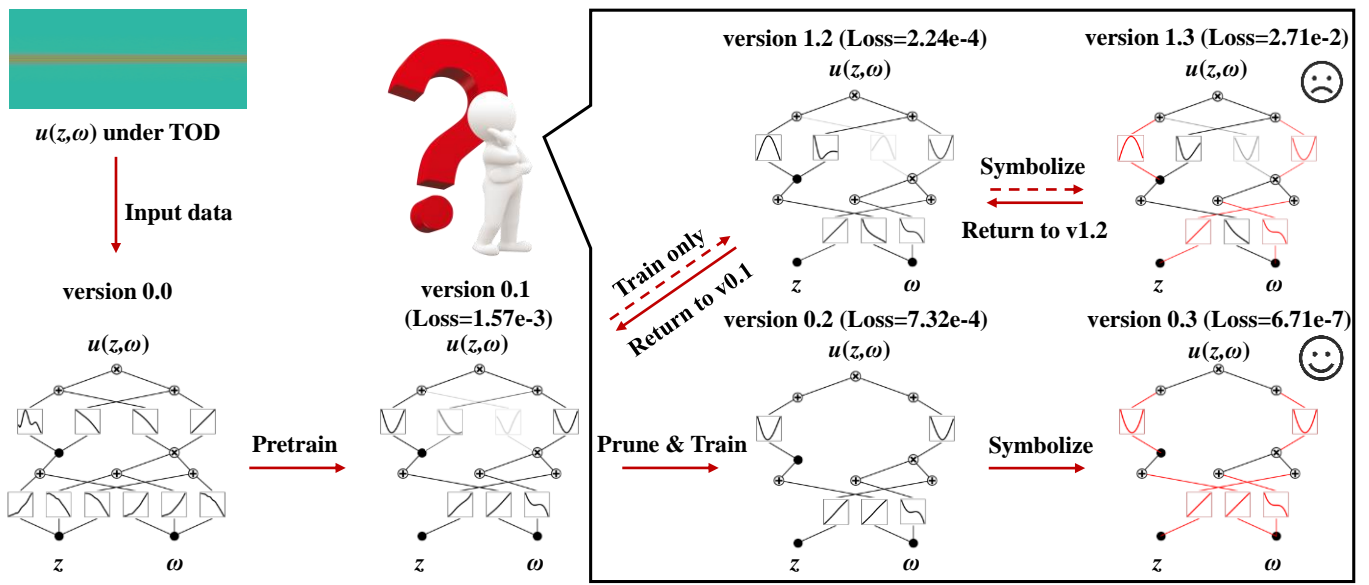

Fig. 16. Hypothesis testing of KANs for discovering $u(z,\omega)$ under the TOD effect, where the typical learning path follows version 0.1 → version 0.2 → version 0.3 and the hypothetical alternative path follows version 0.1 → version 1.2 → version 1.3.

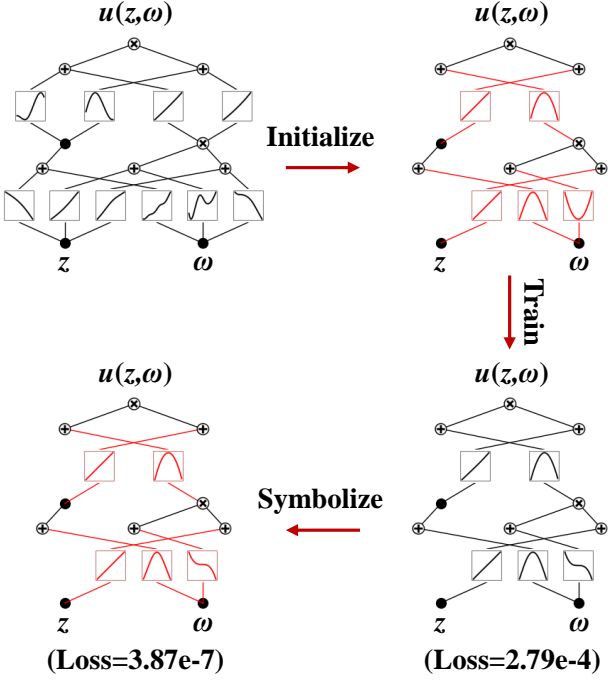

Fig. 17. Transfer learning of KANs for learning $u(z,\omega)$ under the TOD effect from $u(z,\omega)$ under the GVD effect.

### B. Hypothesis Testing of KAN

In practical verification, not every attempt can directly obtain the actual molecular structure followed by the system dynamics. In some attempts, there are often multiple reasonable hypotheses, but most of them may cause KAN to fall into a local optimal solution. In this case, the system dynamics can still be well learned through hypothesis testing, which is another interactive approach adding human's thinking and knowledge. When confronted with multiple plausible hypotheses, we can evaluate each to determine which is the most accurate and suitable. During this process, checkpoint mechanism will save model version whenever network changes (e.g., training, pruning, and symbolizing) are made, so that help us quickly perform hypothesis testing.

Since learning the TOD effect in analytical dynamics is more likely to fail according to Fig. 10, here we still use the case of learning $u(z, \omega)$ under the TOD effect to illustrate the strategy of hypothesis testing. As shown in Fig. 16, the training data is input into the initial constructed KAN, and it is regarded as version 0.0. When it is pretrained to convergence, the loss is $1.57 \times 10^{-3}$, and the model is saved as version 0.1. Since the weights of second and third edges in the second KAN layer are not very small, there are two possible paths: pruning and training or continuing to train the model. Without prior knowledge, it is impossible to determine which path is better, while hypothesis testing allows both paths to be tried in parallel. It can be seen that the losses of both paths converge to the order of $10^{-4}$, and the train-only model (version 1.2) achieves a lower loss value. However, some activation functions of the train-only model do not approximate the basic symbols, which makes them difficult to be symbolized, and the error will increase due to improper symbolization (version 1.3). On the contrary, the model (version 0.3) after pruning, training, and symbolization achieves machine-precision accuracy, illustrating this path is more suitable. With the help of checkpoint mechanism, the train-only model could quickly return to version 0.1 without reloading or reconstructing. In addition, hypothesis testing can also help in scenarios involving redundant layers. For redundant layers, all activation functions of this layer will be learned as the identity function, which guides us to design a KAN structure with fewer layers. By trying different paths, it is able to determine whether the number of layers initially set is appropriate based on subsequent results.

### C. Transfer Learning of KAN

In a fiber-optic system, when the type of fiber or the operating wavelength changes, its corresponding system parameters such as attenuation, dispersion, and nonlinearity will also change. Under different parameters, the system will exhibit similar dynamics, and only simple modification is required. Unfortunately, it is not the case with MLPs, as they usually face the classic catastrophic forgetting problem in DL [49], and retraining is usually required for transfer learning tasks. Although a generalized model can be established by learning data under different parameters, it will inevitably incur high data acquisition cost and introduce generalization error. In contrast, since the solution of system dynamics is decomposed in KANs, such transfer task can be easily accomplished by KANs. Herein, we consider the case where the operating wavelength is converted to the zero-dispersion wavelength. In this scenario, the TOD effect replaces the GVD effect as the dominant chromatic dispersion effect. Since the nature of two effects is similar, their dynamical solutions share the same molecular structure, and the KAN model for TOD dynamics can be transferred from the KAN model for GVD dynamics. First, a KAN model can be initialized from the trained model for $u(z,\omega)$ under the GVD effect, as shown in Fig. 17. Then, the new dataset under the TOD effect is adopted for further training. After only 100 iterations, KAN is able to migrate from the structure of $u(z,\omega)$ under the GVD effect to that under the TOD effect. Since only one activation function changes, no pruning and additional training are required. After the necessary symbolization, $u(z,\omega)$ under the TOD effect is quickly learned. Moreover, in addition to "trying the simplest thing first" and hypothesis testing, the idea of transfer learning can also help us determine the appropriate KAN structure, which greatly reduce the difficulty of structure optimization for similar applications. Therefore, KAN's transfer learning makes the discovery of similar system dynamics more convenient and faster.

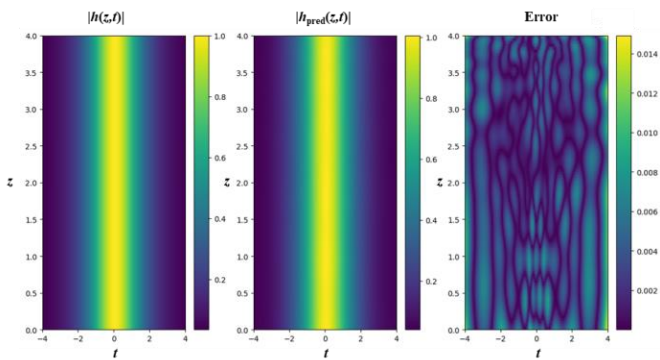

Fig. 18. Comparison between the analytical solution and generated solution obtained by efficient KAN, as well as their absolute errors.

### D. Comparison with Related Algorithms

As one of the physically interpretable structures, KAN is inevitably compared with advanced physics-informed models such as PINN. In scientific discovery problems, comparing Fig. 6 with the PINN learning results of fundamental soliton in Fig. 7 of [19], the accuracy of KAN is several orders of magnitude higher than that of PINN, which illustrates that KAN has extremely high accuracy in scientific discovery problems by directly discovering the analytical solutions. In dynamic characterization problems, KAN can achieve comparable accuracy to PINN with a smaller architecture and fewer trainable parameters. Similarly, comparing Fig. 8(b) with Fig. 6 of [19], KAN shows an error of the same order of magnitude as PINN, and even its maximum error is smaller. This result shows that even in the problem of dynamic representation, KAN still has an accuracy advantage over PINN, and KAN achieves the similar performance to PINN with fewer trainable parameters. However, it should be noted that PINNs are able to realize accurate dynamic modeling without the data in the solving domain, which is of great significance for scenarios with high data acquisition costs.

In addition to the accuracy, the interpretability of KAN is also an important reason why it has attracted widespread attention. Although many variations have been proposed to improve accuracy, computational efficiency, and scenario adaptability of vanilla KAN. However, one of the charms of vanilla KAN is that it can visualize and symbolize each trainable activation function, which is a feature that most variants do not have. To better illustrate the interpretability of the vanilla KAN, here we construct a single-layer efficient KAN with 4 nodes in the KAN layer, which is the same structure as the vanilla KAN for learning the fundamental soliton. Since efficient KAN applies simpler Chebyshev polynomials instead of B-splines as activation functions, it has faster training speed. The results of using efficient KAN to learn fundamental soliton dynamics are shown in Fig. 18. Although the evolution of the fundamental soliton is well predicted, the error cannot be further reduced because the efficient KAN does not include key functions such as visualization of activation functions, pruning, and symbolization. Therefore, the variants of KAN are not interpretable enough to accurately obtain analytical solutions in scientific discovery problems. However, when interpretability is not the primary concern, efficient KANs can quickly characterize the nonlinear dynamics of a physical system.

Moreover, as an important class of methods in the field of scientific discovery, symbolic regression identifies the underlying mathematical expression that best describes the relationship between independent variables and their dependent variable based on genetic algorithm. By combining basic operations (e.g., add, minus, multiply, divide) and variables, symbolic regression explores a vast space of potential models, balancing accuracy and simplicity via fitness metrics and model complexity. However, symbolic regression will face challenges in complex analytical expressions, where more generations are required and it is even impossible to discover the analytical solutions. For the scientific discovery of TOD effect, it requires a long symbolic expression to express its analytical solutions in symbolic regression. Although it can be finally discovered by adjusting the hyperparameters in symbolic regression, rich parameter adjustment experience and more testing time are usually required. In contrast, KAN can greatly improve the success rate of scientific discovery by flexibly adjusting its activation functions to learn complex function terms. Nevertheless, improved algorithms represented by AI Feynman have greatly improved the accuracy and efficiency of symbolic regression [50].

## VI. CONCLUSION

In this paper, a KAN-based AI4S framework FiberKAN was proposed for discovering and characterizing the underlying dynamics in fiber-optic systems. Different from the classic MLPs, the activation functions in KANs are trainable, transparent, and interpretable, enabling the network to better explain and learn the dynamics of complex systems. Results showed that KANs can well learn the corresponding analytical solutions under power attenuation, chromatic dispersion, and the combined effects of GVD and SPM, as well as the implicit equations satisfied under the SS effect. For the non-analytical IRS dynamics, KAN exhibited superior nonlinear learning ability compared to MLPs with equivalent network scale. From the perspective of practical applications, the effectiveness, computational cost, and noise resistance of FiberKAN were comprehensively evaluated. In addition, interactive approaches such as prior knowledge embedding and hypothesis testing were also analyzed to better unleash the potential of KANs in scientific discovery. Furthermore, KAN also showcased the powerful transfer learning capability in similar systems. Compared with related algorithms including PINN, variant algorithms, and symbolic regression, FiberKAN has advantages in accuracy, interpretability, and implementation difficulty. As a promising network structure, KAN is expected to be an alternative to MLP that realizes scientific discovery and dynamic characterization of physical systems including fiber-optic systems, thereby promoting the progress of AI4S.